\documentclass[%
reprint,
 amsmath,amssymb,
 aps,
prb,
]{revtex4-1}
\usepackage[latin1]{inputenc}
\usepackage{mathtools}
\usepackage{slashed}
\usepackage{physics}
\usepackage{array}
\usepackage{amsmath}
\usepackage{color}
\usepackage{amsfonts}
\usepackage{amssymb}
\usepackage{graphicx}
\usepackage{amssymb,epsfig,subfigure}
\usepackage{hyperref}
\usepackage{comment}
\usepackage{tabularx}
\usepackage{bm}
\usepackage{euscript}
\usepackage{graphicx}
\usepackage{color}
\usepackage [english]{babel}
\usepackage [autostyle, english = american]{csquotes}
\MakeOuterQuote{"}
\usepackage{amsfonts}
\usepackage{exscale}
\usepackage{amsbsy}
\usepackage{subfigure}
\usepackage{textcomp}
\usepackage{comment}
\usepackage{hyperref}
\usepackage{slashed}
\usepackage{tabularx}
\usepackage{euscript}
\usepackage{graphicx}
\usepackage{color}
\usepackage{amsfonts}
\usepackage{exscale}
\usepackage{amsbsy}
\usepackage{subfigure}
\usepackage{textcomp}
\usepackage{comment}
\usepackage{hyperref}
\usepackage{bm}
\usepackage{wrapfig}
\usepackage[font=footnotesize,labelfont=bf]{caption}

\def\ba#1\ea{\begin{align}#1\end{align}}
\def\bg#1\eg{\begin{gather}#1\end{gather}}
\def\bm#1\em{\begin{multline}#1\end{multline}}
\def\bmd#1\emd{\begin{multlined}#1\end{multlined}}

\newcommand{\beq}{\begin{eqnarray}}
\newcommand{\eeq}{\end{eqnarray}}
\begin{document}

\def\ppnumber{\vbox{\baselineskip14pt
}}

\def\ppdate{
} \date{\today}

\title{\bf Twisted Kitaev Bilayers and the Moir\'e Ising Model}
\author{Julian May-Mann}
\affiliation{ \it Department of Physics and Institute for Condensed Matter Theory,\\  \it University of Illinois at Urbana-Champaign, \\  \it 1110 West Green Street, Urbana, Illinois 61801-3080, USA}
\author{Taylor L. Hughes}
\affiliation{ \it Department of Physics and Institute for Condensed Matter Theory,\\  \it University of Illinois at Urbana-Champaign, \\  \it 1110 West Green Street, Urbana, Illinois 61801-3080, USA}

\begin{abstract}

In recent years, there have been numerous examples of twisted bilayer systems that host remarkable physical properties that are not found in their untwisted counterparts. Motivated by this, we study the properties of twisted bilayers of the Kitaev honeycomb model in the Abelian spin liquid phase. We show that for strong, short-ranged, interlayer interactions, a super-lattice of non-Abelian defects forms in the twisted bilayer system. These non-Abelian defects are wormhole-like genons that allow anyons from one layer to tunnel to the other layer. We find that when a magnetic field is applied to the system, the low energy dynamics of the twisted bilayer system can be mapped onto four quantum Ising models arising from the degrees of freedom localized on the genon defects. At small twist angles, the Ising models are in a trivial paramagnetic phase, and at large twist angles, they are in a ferromagnetic phase. 
\end{abstract}

\maketitle

\bigskip
\newpage



\section{Introduction}

\begin{figure}
\centering
\includegraphics[width=\linewidth*3/4]{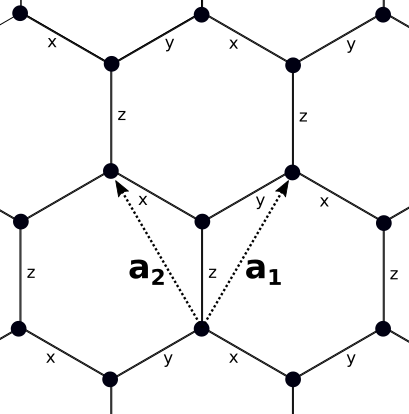}
\caption{The honeycomb lattice with primitive vectors $\textbf{a}_1$ and $\textbf{a}_2$, and $x,y,$ and $z$-oriented links.}
\label{fig:kLinks}
\end{figure}

A quantum spin liquid is an exotic state of strongly correlated spins, that does not break any symmetries\cite{savary2016}. The study of spin liquids began in 1973 with Anderson's work on the resonating valence bond state\cite{anderson1973}. Since then, spin liquids have been an active area of theoretical and experimental research. In particular, it has been shown that topological order plays a major role in the structure of these systems\cite{kivelson1987, rokhsar1988, wen1991}. A well known example of a topologically ordered spin liquid appears in the spin-$1/2$ Kitaev model\cite{kitaev2006}. This model is defined on a honeycomb lattice with one spin-$1/2$ at each vertex of the lattice. The Hamiltonian is given by
\beq
\nonumber H_{\text{Kitaev}} = &-&J_x\sum_{\langle r,r'\rangle \in x}\sigma^x_r \sigma^x_{r'} - J_y\sum_{\langle r,r'\rangle \in y}\sigma^y_r \sigma^y_{r'} \\&-& J_z\sum_{\langle r,r'\rangle \in z}\sigma^z_r \sigma^z_{r'},
\label{eq:kModel}
\eeq
where in the first sum, $\langle r,r'\rangle \in x$ indicates that the sites $r$ and $r'$ are neighboring sites that are connected by an $x$-oriented link and similarly for $y$ and $z$ (see Fig. \ref{fig:kLinks}). The Kitaev model can be exactly solved by decomposing each spin into 4 Majorana fermions. Depending on the relative strengths of the interactions, $J_x$, $J_y$, and $J_z$, there are two phases of Eq. \ref{eq:kModel}. First, there is a gapped $\mathbb{Z}_2$ Abelian topologically ordered spin liquid phase. Second, there is a gapless phase of (effectively) free Majorana fermions. If one adds  an external magnetic field, the free Majorana fermions can acquire a gap such that the system enters a third phase that has non-Abelian topological order.

Recently, it has been proposed that $\alpha$-RuCl$_3$ can realize the non-Abelian spin liquid state of the Kitaev model \cite{plumb2014, sandilands2015, sears2015, kim2015, majumder2015, banerjee2016, banerjee2017}. Crystals of $\alpha$-RuCl$_3$ are made up of layers of Ru and Cl atoms, which form 2D honeycomb lattices. Since $\alpha$-RuCl$_3$ has this honeycomb structure, as well as strong spin-orbit coupling, it is an appealing candidate to realize the Kitaev model. Other candidate materials include Na$_2$IrOs$_{3}$, $(\alpha,\beta,\gamma)$-Li$_2$IrO$_3$, and H$_3$LiIr$_2$O$_6$\cite{jackeli2009,modic2014,takayama2015,bette2017,takagi2019}.

In a completely disparate physical context, twisted honeycomb bilayer systems have also become an active area of research. When one layer of a bilayer honeycomb lattice is twisted with respect to the other layer, a moir\'e pattern forms, and a super-lattice structure can emerge when this moir\'e pattern is commensurate with the original honeycomb lattice. Research on twisted bilayer systems has been primarily centered on twisted bilayer graphene\cite{morell2010, mele2010, bistritzer2011}. In particular, unconventional superconductivity has been observed in graphene bilayers that have been twisted to the "magic angle" of $\theta \approx 1.1^{o}$\cite{cao2018,cao2018b}. Understanding magic angle twisted bilayer graphene remains an exciting open theoretical and experimental question\cite{padhi2018,zou2018}.

Motivated by these advances in the studies of spin liquids and twisted bilayer systems, we will consider \emph{commensurate} twisted bilayers of the Kitaev honeycomb model where both layers are tuned into the Abelian spin liquid phase. Our analysis will focus on the effects of strong, short-ranged, interlayer couplings. By constructing an effective Hamiltonian, we show that in the regime of strong interlayer couplings, a super-lattice of non-Abelian defects forms in the twisted bilayer system, despite the underlying phase being Abelian in nature. These non-Abelian defects can be thought of as \textit{wormhole}-like "genons"\cite{barkeshli2013}, which allow fractionalized quasiparticles (anyons) to pass from one layer to the other. If no magnetic field is present, these defects result in a ground state degeneracy that is exponential in the size of the system. This ground state degeneracy corresponds to the existence of non-trivial zero energy loop operators that enclose or pass through the wormhole defects. When a weak magnetic field is applied to the system, the ground state degeneracy is split. To understand this splitting we construct an effective Hamiltonian for the aforementioned loop operators, and we show that the resulting low energy dynamics of the twisted bilayer system can be mapped onto an effective spin model consisting of four decoupled 2D quantum Ising models. These quantum Ising models have position depending Ising couplings as well as position dependent transverse magnetic field terms. Furthermore, we show that the resulting phases of the Ising models depend on the twist angle of the bilayers. At small twist angles, the Ising models favor a trivial paramagnetic phase, while at large twist angles, a ferromagnetic phase of the effective spin degrees of freedom is favored.

This paper is organized as follows. In Section \ref{Review} we review the Abelian phase of the Kitaev honeycomb model. In Section \ref{DefectLattice} we present the bilayer model and analyze the twisted Kitaev bilayers. We show that at strong coupling, the system hosts a lattice of non-Abelian defects. In Section \ref{MIsingModel} we analyze the dynamics of the defect lattice in the presence of a magnetic field, and show that the low energy physics is well described by an effective spin model of four quantum Ising models. Furthermore, we show that the phases of the Ising models depend on the twist angle between the bilayers. We conclude our results in Section \ref{Conclusion} and discuss possible extensions. 

\section{Review: Abelian Phase of the Kitaev Model}
\label{Review}
\begin{figure}
\centering
\includegraphics[width=\linewidth/2]{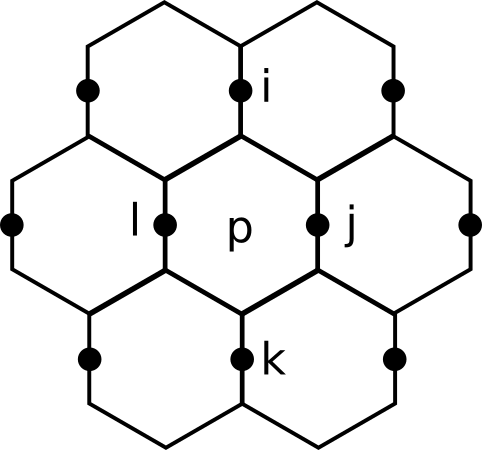}
\caption{The locations of the effective spins ($i$, $j$, $k$, and $l$) that make up the plaquette $p$.}
\label{fig:kEffDia}
\end{figure}

\begin{figure}
\centering
\includegraphics[width=\linewidth*3/4]{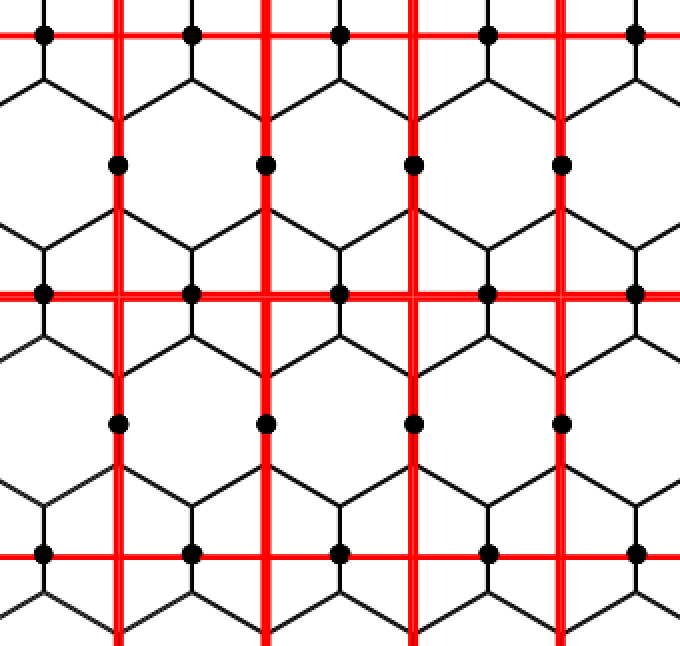}
\caption{The original honeycomb lattice (black), and the new square lattice (red). }
\label{fig:kEffLat}
\end{figure}

To begin, we will review the Abelian phase of the Kitaev model (Eq. \ref{eq:kModel}). In our analysis we will take $J_x$, $J_y$, and $J_z$ to all be positive. For a full analysis, see Kitaev's seminal paper on the model\cite{kitaev2006}. It is known that the Kitaev model is in a $\mathbb{Z}_2$ Abelian topologically ordered phase if $J_z >   J_x+J_y$. This can be shown explicitly by considering the case where $J_z \gg   J_x,J_y,$ and then applying perturbation theory by treating the $J_x$ and $J_y$ terms as perturbations. When $J_x=J_y=0$, the spins that are connected by a $z$-oriented links are aligned, i.e., if $\langle r,r'\rangle \in z$, then $\langle \sigma^z_r \rangle = \langle \sigma^z_{r'} \rangle$. Because of this, we can consider each $z$-oriented link to be a single spin-$1/2$. Let us consider a single $z$-oriented link, $i$, that connects two sites $r$ and $r'$. The Pauli matrices for the effective spin-$1/2$ located at $i$ are:
\begin{equation}
\begin{split}
    &\bar{\sigma}^z_i \equiv \sigma^z_r = \sigma^z_{r'},\\ &\bar{\sigma}^x_i \equiv \sigma^x_{r}\sigma^x_{r'} = \sigma^y_r\sigma^y_{r'},\\ &\bar{\sigma}^y_{i} \equiv \sigma^x_r\sigma^y_{r'} = \sigma^y_{r}\sigma^x_{r'} .
\end{split}
\label{eq:PMDef1}
\end{equation}
To leading order in perturbation theory, the effective Hamiltonian for these spin-$1/2$s is
\beq
H_{A} = -J_{\text{eff}} \sum_{ p} \bar{\sigma}^z_i\bar{\sigma}^y_j\bar{\sigma}^z_k\bar{\sigma}^y_l,
\label{eq:kEff}
\eeq 
where $p$ is the plaquette formed by the effective spins at sites $i$, $j$, $k$, and $l$ (see Fig. \ref{fig:kEffDia}), and to leading order $J_{\text{eff}} = \frac{J_x^2J_y^2}{16 J_z^3}$. To bring Eq. \ref{eq:kEff} into a more familiar form, we will define a new square lattice where the effective spins lie on the links of the square lattice (see Fig. \ref{fig:kEffLat}). If we appropriately rotate the effective spins of the square lattice, the effective Hamiltonian becomes,
\beq
H_{TC} = -J_{\text{eff}}\sum_{v}\prod_{l\in v} \bar{\sigma}^x_l  -J_{\text{eff}}\sum_{p}\prod_{l\in p} \bar{\sigma}^z_l,
\label{eq:TorHam}
\eeq
where the two sums are over the vertices $v$ and plaquettes $p$ of the square lattice, and the products are over the links that make up a given vertex or plaquette. Eq. \ref{eq:TorHam} is the well known Hamiltonian for the toric code\cite{kitaev2003}. If we return to the original honeycomb lattice, we see that vertices and plaquettes of the square lattice in Fig. \ref{fig:kEffLat} correspond to alternating rows of plaquettes in the original honeycomb lattice. 

We will now include a quick summary of the features of the toric code model\cite{kitaev2003}. All terms in Eq. \ref{eq:TorHam} commute with each other, and square to unity. Thus the ground state is achieved in the quantum number sector that minimizes each term. Since $J_{\text{eff}} > 0$, this occurs when $\langle \prod_{l\in v} \bar{\sigma}^x_l \rangle = \langle \prod_{l\in p} \bar{\sigma}^z_l \rangle = 1$ for all vertices $v$ and plaquettes $p$ in the system. The spectrum of Eq. \ref{eq:TorHam} is gapped, and when the system is defined on a system of genus $g$, there are $4^g$ degenerate ground states. 

The quasiparticles of Eq. \ref{eq:TorHam} are typically referred to as "anyons" due to their unusual statistical properties. There are three types of non-trivial toric code anyons: $e$ anyons, which correspond to a vertex $v$ where $\langle \prod_{l\in v} \bar{\sigma}^x_l \rangle = -1$, $m$ anyons, which correspond to a plaquette $p$ where $\langle \prod_{l\in p} \bar{\sigma}^z_l \rangle = -1$, and $\psi$ anyons, which are the fusion of an $e$ and $m$ anyon. The $e$ and $m$ anyons are both bosons, and $\psi$ is a fermion. The statistical angle between the $e$ and $m$ is $\pi$. The fusion rules of the anyons are $e\times e = 1$, $m \times m = 1$ and $e\times m = \psi$, and all toric code anyons are their own anti-particles. 

When a pair of anyons is created, they are connected by a string operator. Annihilating this pair of anyons causes the string to close, forming a loop that we will refer to as an anyon loop. The operators that create these anyon loops commute with the Hamiltonian in Eq. \ref{eq:TorHam}, and in the language of gauge theories, the anyon loops are the gauge invariant Wilson loops of the theory. There are two types of anyon loops. First there are contractible anyon loops that can be smoothly deformed into a single point. All contractible anyon loops commute with all other contractible anyon loops. Indeed, the toric code ground state can be interpreted as a condensate of these contractible anyon loops\cite{levin2005}. Second, on a lattice with genus greater than zero, there are also non-contractible anyon loops that cannot be smoothly deformed to a point. In general, non-contractible anyon loops do not commute with each other. The algebra of non-contractible anyon loops leads to the aforementioned ground state degeneracy on surfaces with genus greater than zero. 

Dynamics for the anyons can be added to Eq. \ref{eq:TorHam} by including a magnetic field term:
\beq
H_{\text{mag}} = -\sum_l[  h_x \bar{\sigma}^x_l + h_y \bar{\sigma}^y_l + h_z \bar{\sigma}^z_l].
\label{eq:AnyonHopping}
\eeq
Comparing Eqs. \ref{eq:TorHam} and \ref{eq:AnyonHopping}, we note that the $h_x$ term creates (or removes) a pair of excited states at neighboring plaquettes, i.e., creates a pair of $m$ anyons. Similarly the $h_z$ term creates a pair of $e$ anyons, and the $h_y$ term creates a pair of $\psi$ anyons. By acting multiple times with these operators, anyons can be moved around the system. Provided that the strength of the magnetic field is significantly smaller than the bulk gap of the system, the effects of Eq. \ref{eq:AnyonHopping} can be studied perturbatively around the ground state of Eq. \ref{eq:TorHam} in inverse powers of the system gap.

\section{Twisted Bilayers and the Defect Lattice}
\label{DefectLattice}
The staring point for our analysis will be two copies of the Kitaev model on a honeycomb lattice. We will label the Pauli matrices acting on the top layer as $\sigma_{\uparrow}$ and the Pauli matrices acting on the bottom layer as $\sigma_{\downarrow}$. The Hamiltonian is given by: 
\beq
\nonumber H_{\text{bilayer}} = &-&J_x\sum_{\langle r,r'\rangle \in x}\sigma^x_{\uparrow,r} \sigma^x_{\uparrow,r'} - J_y\sum_{\langle r,r'\rangle \in y}\sigma^y_{\uparrow,r} \sigma^y_{\uparrow,r'} \\&-& J_z\sum_{\langle r,r'\rangle \in z}\sigma^z_{\uparrow,r} \sigma^z_{\uparrow,r'} + (\sigma_\uparrow \leftrightarrow \sigma_\downarrow ).
\label{eq:kModel2}
\eeq
We are interested in the Abelian phase of the model where $J_z \gg J_x,J_y$ for both layers. The interlayer couplings will take the form of a short-ranged, antiferromagnetic, Heisenberg interaction between the spins of the two layers:
\beq
H_{\text{inter}} = K \sum_{r \in C} \vec{\sigma}_{\uparrow,r}\cdot \vec{\sigma}_{\downarrow,r}
\label{eq:sLink}
\eeq
where $K$ is positive and $C$ is the set of all coincident vertices of the two layers. Since the magnetic interaction is short-ranged, we will assume that spins only at the coincident sites of the bilayer system ($r \in C$) are coupled. If the bilayer is untwisted, then all sites of the bilayer lattice will be coincident. If the bilayer is twisted commensurately only certain sites will be coincident. This will be discussed in more detail in Sec. \ref{SSTwistAngles}. If $K$ is much smaller than the relevant energy scales of the model ($K \ll J_\alpha$, $\alpha = x,y,z$), then Eq. \ref{eq:sLink} can be treated as a perturbation to Eq. \ref{eq:kModel2}. However, since each layer in Eq. \ref{eq:kModel2} is gapped, this perturbation is irrelevant at weak coupling. We thereby do not expect such a perturbation to change any universal features associated with the decoupled bilayers. At strong coupling ($K\gg  J_\alpha$), it is clear that to zeroth order, the spins at site $r$ will form a singlet. Higher order corrections are more intricate and will be discussed further below.

Now we shall analyze the coupled bilayer systems in more detail. As noted before, the interlayer interactions are irrelevant at weak coupling, and, as such, we will focus on the limit of strong interlayer coupling. 

\subsection{Untwisted Bilayers}
\label{SSUntwisted}
Before considering twisted bilayers, we will first consider untwisted bilayers (also referred to as AA stacked bilayers). When the bilayer is untwisted, all sites of the bilayer system are coincident, and the interlayer coupling is given by
\beq
H_{\text{inter}} = \sum_{r}K \vec{\sigma}_{\uparrow,r}\cdot \vec{\sigma}_{\downarrow,r}
\eeq
where the sum is over \textit{all} sites of the honeycomb lattice bilayer. As before, at weak coupling, $K \ll J_\alpha$, the interlayer coupling is irrelevant, due to the bulk gap, and the system resembles two decoupled copies of toric code. In the strong coupling limit, $K \gg J_\alpha$, the two spins at each site form a singlet, and the system is a trivial magnet with interlayer dimerization. The intermediate coupling regime may prove to be interesting, but it is beyond the scope of this paper. In particular, determining the nature of the phase transition that connects the weak and strong coupling regimes may be a topic for future work. A related analysis of untwisted Kitaev bilayers in the non-Abelian phase has been done by Seifert et. al.\cite{seifert2018}. Similarly, they find that at weak interlayer coupling the system is a non-Abelian spin liquid, and at strong coupling the system forms trivial interlayer singlets. 

\subsection{Commensurate Twist Angles}
\label{SSTwistAngles}
Having established that the untwisted Kitaev bilayer system is trivial at strong coupling, we will now turn our attention to the more interesting situation where the bilayers are twisted. In general, there are two types of twist angles: commensurate or incommensurate with the underlying honeycomb lattice. When the twist angle is commensurate, the twisted bilayer system has a super-lattice structure, and translational symmetry with respect to this super-lattice. When the twist angle is incommensurate, there is no such super-lattice structure or translational symmetry. In this study, we are interested in this super-lattice, and so we will restrict our attention to commensurate twist angles. 

\begin{figure}
\centering
\includegraphics[width = \linewidth*3/7]{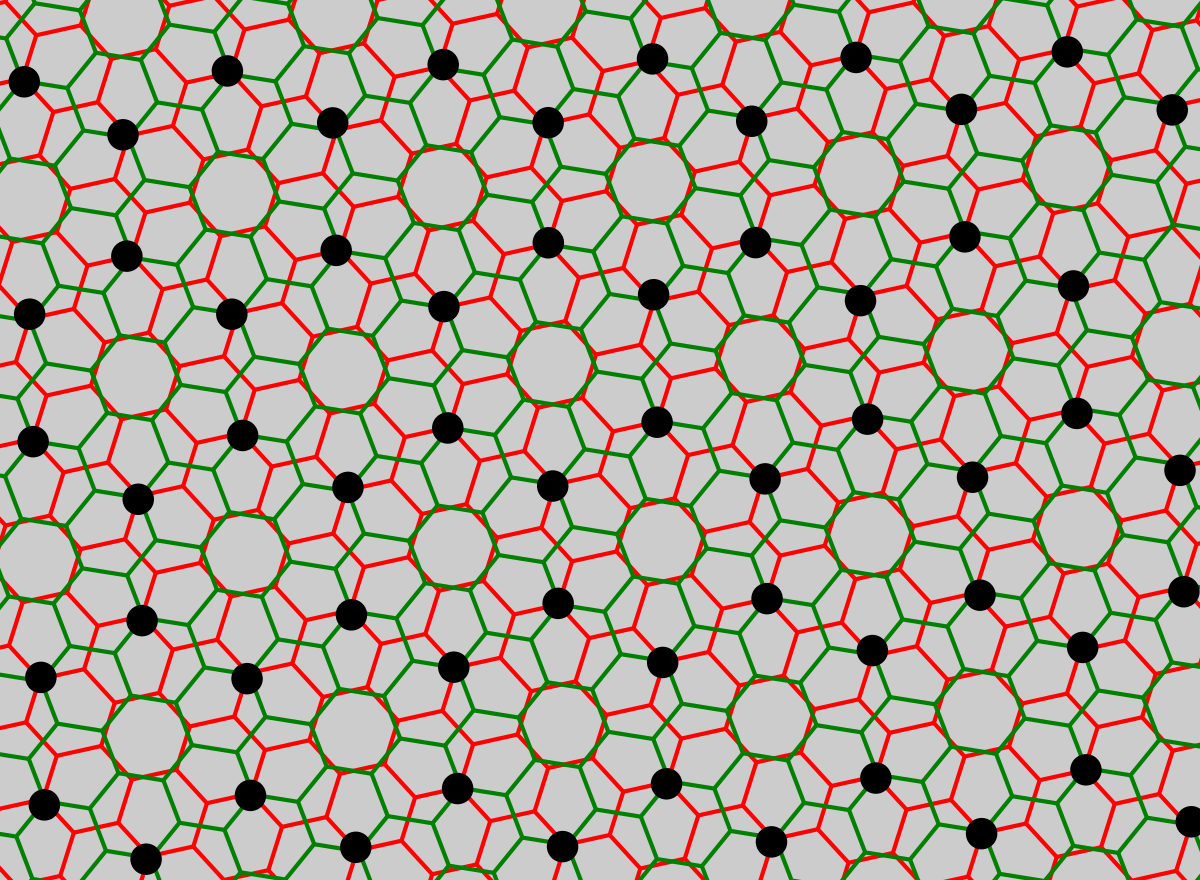}
\includegraphics[width = \linewidth*3/7]{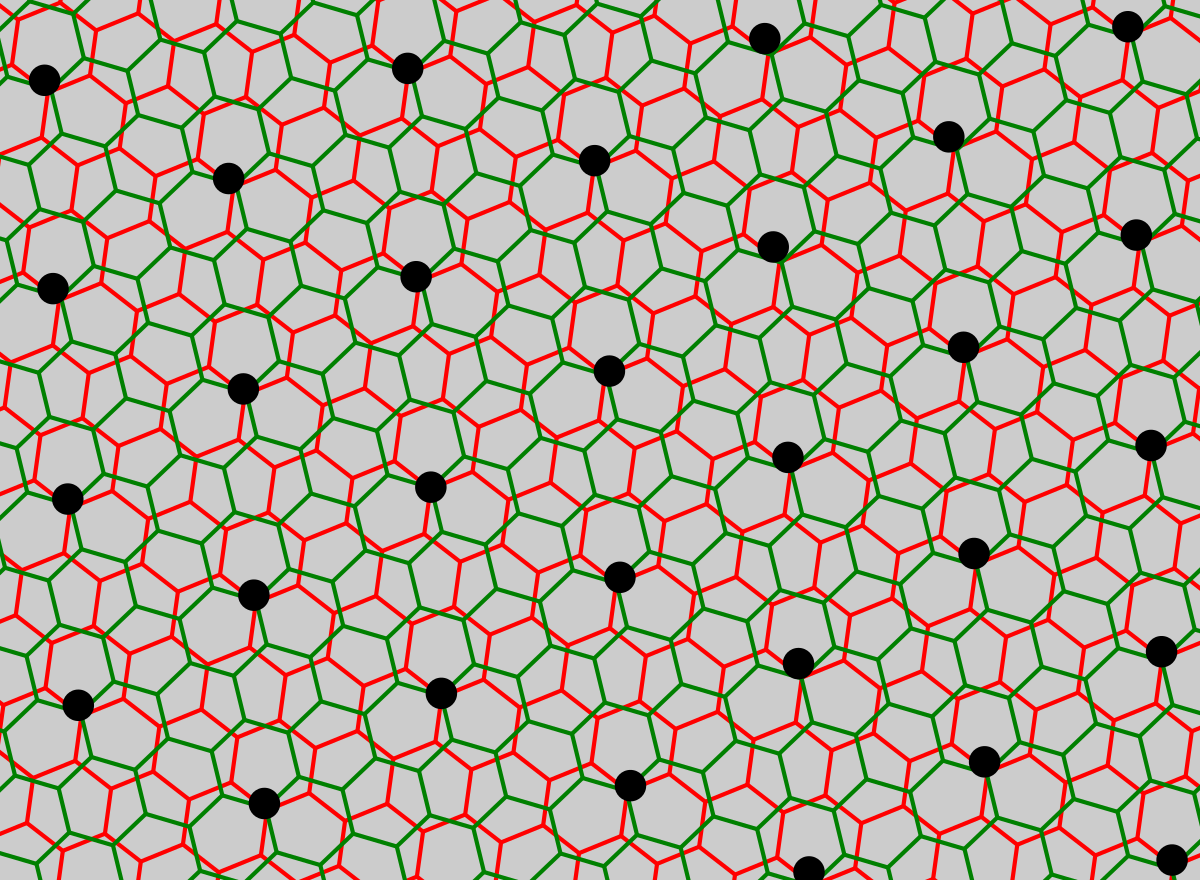}
\caption{The twisted honeycomb lattice bilayer for $\theta(1,3) \approx 38.21^o $ (left) and $\theta(1,1) \approx 21.79^o $ (right). The coincident lattice points are shown in black. These points form either a honeycomb lattice (left) or triangular lattice (right).}
\label{fig:TwistLat}
\end{figure}

The set of commensurate twist angles is given by the equation \cite{shallcross2010,mele2010}
\beq
\cos(\theta(m,r))=\frac{3m^2+3mr+r^2/2}{3m^2+3mr+r^2},
\label{eq:twistEq}
\eeq
where $m$ and $r$ are co-prime integers, and $0<\theta <\pi / 3$. At these twist angles, only certain sites of the bilayer system will be coincident. These coincident sites form a super-lattice (see Fig. \ref{fig:TwistLat}). The primitive vectors of this super-lattice, $ \textbf{t}_1$ and $ \textbf{t}_2$, are:

i. If $\text{gcd}(r,3)=1$,
\beq
\begin{bmatrix}
\textbf{t}_1 \\ \textbf{t}_2
\end{bmatrix} = \begin{bmatrix}
m & m+r \\ -m-r & 2m+r 
\end{bmatrix} \begin{bmatrix}
\textbf{a}_1 \\ \textbf{a}_2
\end{bmatrix},
\label{eq:PV1}
\eeq

ii. If $\text{gcd}(r,3)=3$,
\beq
\begin{bmatrix}
\textbf{t}_1 \\ \textbf{t}_2
\end{bmatrix} = \begin{bmatrix}
m+r/3 & r/3 \\ -r/3 & m+2r/3 
\end{bmatrix} \begin{bmatrix}
\textbf{a}_1 \\ \textbf{a}_2
\end{bmatrix},
\label{eq:PV2}
\eeq
where $ \textbf{a}_1 $ and $ \textbf{a}_2 $ are the primitive vectors of the original honeycomb lattice (see Fig. \ref{fig:kLinks}). When $\text{gcd}(r,3) = 1$, the coincident sites form a triangular lattice, and when $\text{gcd}(r,3) = 3$, the coincident sites form a honeycomb lattice. These two super-lattice patterns are referred to as sub-lattice exchange odd (SE-odd) and sub-lattice exchange even (SE-even) respectively. 

\subsection{Effective Action at a Single Site}
\label{SSSingleSite}
As noted, when the twist angle is commensurate, only certain isolated sites of the bilayer lattice are coincident. Under the assumption that the interlayer couplings are short-ranged, only the pairs of spins at these coincident sites will couple to each other. Intuitively, and as indicated in Eqs. \ref{eq:PV1} and  \ref{eq:PV2}, the distance between the coincident sites of the twisted bilayer system is greater than the original lattice spacing of the honeycomb lattice. Because of this, and the fact that the bulk of the Kitaev bilayer system is gapped, spins at different coincident lattice sites will only weakly interact. For this reason, we will first study a system where only a single pair of spins is coupled. 

\begin{figure}
\centering
\includegraphics[width=\linewidth*3/4]{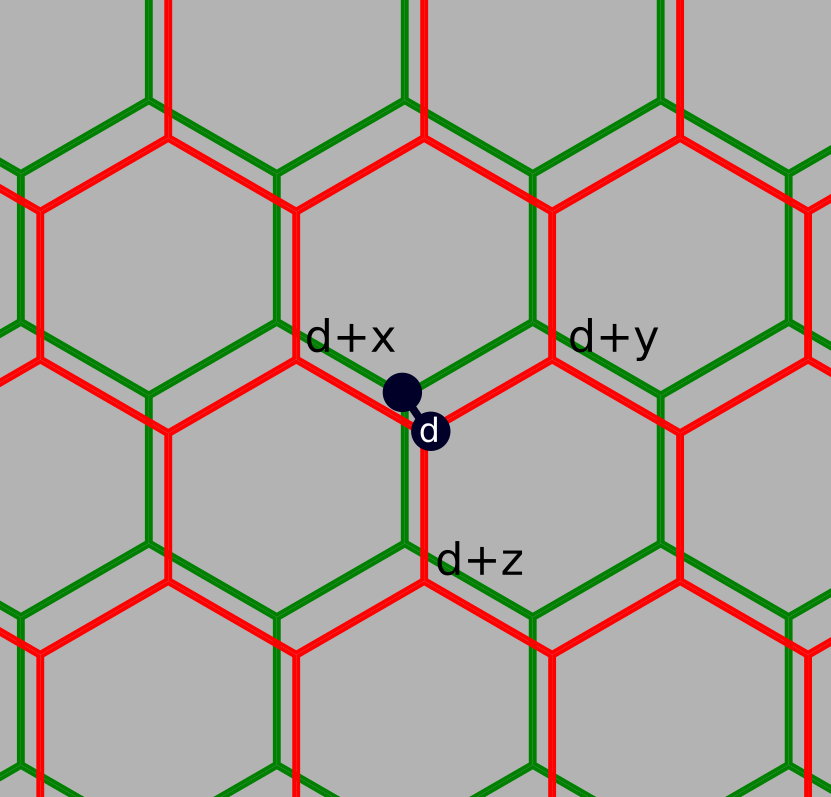}
\caption{The lattice sites that interact with the coupled spins (black) at site $d$.}
\label{fig:coupEffDia}
\end{figure}

For a bilayer system where only a single pair of spins is coupled, the interlattice coupling is given by
\beq
H_{\text{inter}} = K \vec{\sigma}_{\uparrow,d}\cdot \vec{\sigma}_{\downarrow,d},
\label{eq:InterCoup1S}
\eeq 
where $d$ is a \textit{single} coincident site of the bilayer system (see Fig. \ref{fig:coupEffDia}). To study the the strong coupling limit of Eq. \ref{eq:InterCoup1S} ($K \gg J_\alpha$) we will construct an effective Hamiltonian perturbatively in powers of $J_\alpha/K$. Before we do this, we will first review the procedure for constructing an effective Hamiltonian\cite{auerbach2012}. For a Hamiltonian of the form $H = H_0 + V$, we want to find the effective Hamiltonian that  acts \textit{only} on the space of ground states of $H_0$. This is done perturbatively in powers of $V$. Explicitly, the matrix elements of the effective Hamiltonian are given by 
\beq
\nonumber \langle a | H_{\text{eff}} |b \rangle  = E_0 \delta_{a,b} + \langle a |V|b\rangle +  \sum_\gamma \frac{\langle a | V | \gamma \rangle\langle \gamma | V | b \rangle}{E_0 - E_\gamma}+...,\\
\label{eq:EffHamSeries}
\eeq
where $E_0$ is the ground state energy of $H_{0}$, $\ket{a}$ and $\ket{b}$ are ground states of $H_0$, and the sum is over all excited states $\ket{\gamma}$ of $H_0$. 

Since $K\gg J_\alpha$ we will treat the interlayer coupling in Eq. \ref{eq:InterCoup1S} as $H_0,$ and the intralayer couplings in Eq. \ref{eq:kModel2} as $V$. 
In the ground state of Eq. \ref{eq:InterCoup1S}, the two spins at site $d$ will form a singlet, and all other spins in the bilayer system will be free. To leading order, the effective Hamiltonian that acts on the free spins is given by 
\beq
\nonumber H_{\text{eff}} &=& -J_x\sum'_{\langle r,r'\rangle \in x}\sigma^x_{\uparrow,r} \sigma^x_{\uparrow,r'} - J_y\sum'_{\langle r,r'\rangle \in y}\sigma^y_{\uparrow,r} \sigma^y_{\uparrow,r'} \\\nonumber &\phantom{=}& - J_z\sum'_{\langle r,r'\rangle \in z}\sigma^z_{\uparrow,r} \sigma^z_{\uparrow,r'} + (\sigma_{\uparrow} \leftrightarrow \sigma_{\downarrow})\\
\nonumber &\phantom{=}& -J_{t,z} \sigma^z_{\uparrow,d+z}\sigma^z_{\downarrow,d+z} -J_{t,z} \sigma^z_{\uparrow,d+x}\sigma^z_{\downarrow,d+x}\\ &\phantom{=}& -J_{t,y} \sigma^y_{\uparrow,d+y}\sigma^y_{\downarrow,d+y},
\label{eq:effHamFull}
\eeq
where the primed sum is over all sites $r,r' \neq d$, and $d+\alpha$ is the site connected to $d$ by an $\alpha$-oriented link (see Fig. \ref{fig:coupEffDia} for the reference labels). To leading order, $J_{t,\alpha} \sim \frac{J^2_{\alpha}}{K}$. By using a unitary transformation, we can set $J_{t,\alpha}$ to be positive. 

Let us compare the effective Hamiltonian in Eq. \ref{eq:effHamFull} to the original Kitaev Hamiltonian (Eq. \ref{eq:kModel}). Away from the coupled spins, we recover the two original Kitaev models as expected. Near the coupled spins, we note that the $J_{t,z}$ term in Eq. \ref{eq:effHamFull} is what one would have if $\sigma_{\uparrow,d+z}$ and $\sigma_{\downarrow,d+z}$ spins were connected by a $z$-oriented link in the original Kitaev model. Similarly, the $J_{t,x(y)}$ term in Eq. \ref{eq:effHamFull} is what one would have if $\sigma_{\uparrow,d+x(y)}$ and $\sigma_{\downarrow,d+x(y)}$ were connected by an $x(y)$-oriented link. Based on this intuition, we expect that the interactions in Eq. \ref{eq:InterCoup1S} will allow excitations from one layer to tunnel to the other layer. As we shall now show explicitly, this expectation is correct.

\begin{figure}
\centering
\includegraphics[width = \linewidth*5/6]{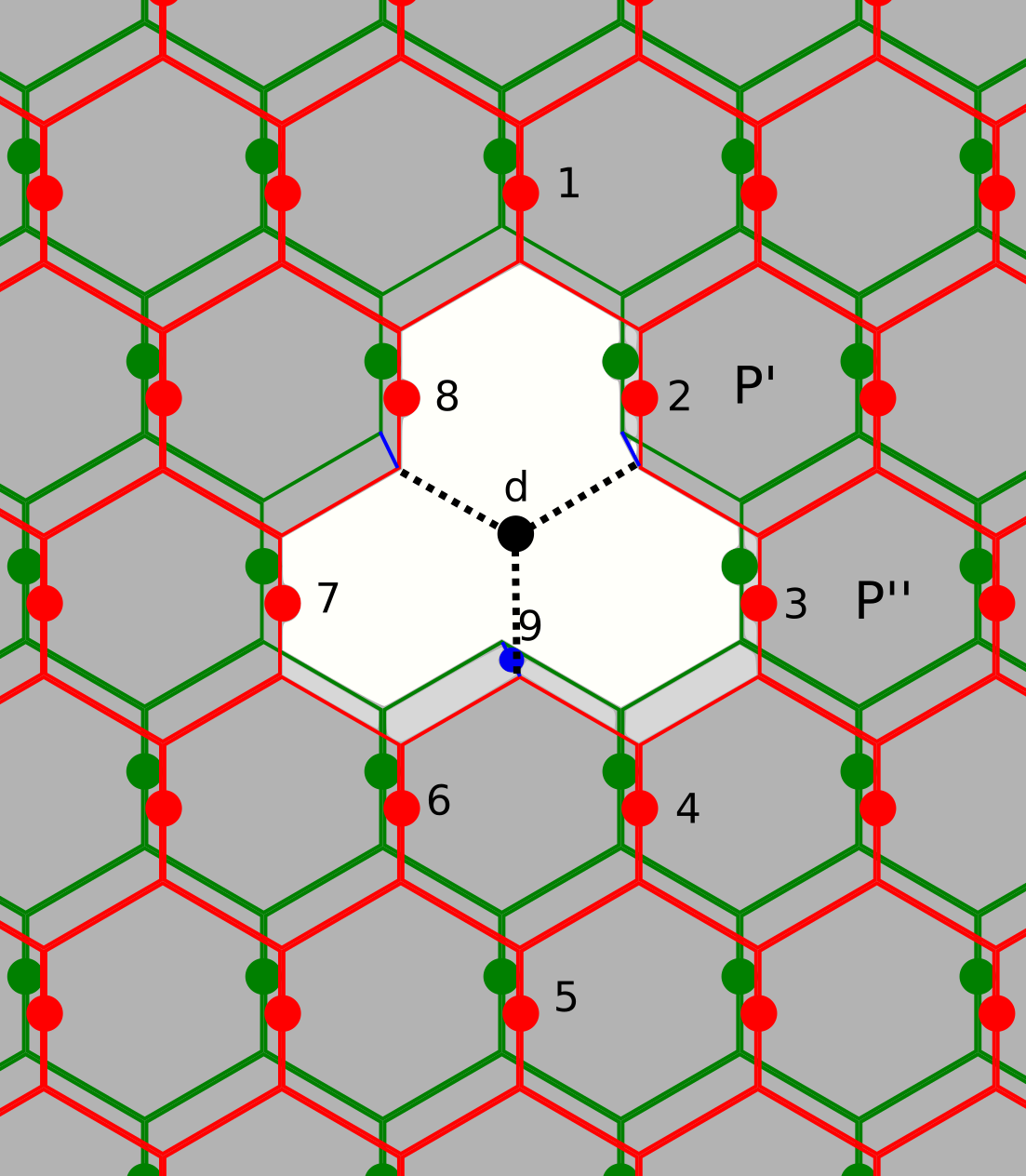}
\caption{The effective spins contributing to the effective Hamiltonian around the defect. The red dots are the effective spins of the top layer and green dots are the effective spins of the bottom layer. The blue dot is the effective spin that is shared between the two layers at $d+z$.}
\label{fig:ATwistEffDia}
\end{figure}

In the Abelian phase, when $J_{z} \gg J_{x},J_y$, we can analyze the effects of the interlayer couplings on the rest of the system by constructing a second effective Hamiltonian from the first effective Hamiltonian Eq. \ref{eq:effHamFull}. For this second effective Hamiltonian we will treat the $J_x$, $J_y$, $J_{t,x}$ and $J_{t,y}$ terms as perturbations. When $J_x=J_y = J_{t,x} = J_{t,y} = 0$, Eq. \ref{eq:effHamFull} becomes 
\beq
\nonumber H_{\text{eff}} = &-& J_z\sum'_{\langle r,r'\rangle \in z}(\sigma^z_{\uparrow,r} \sigma^z_{\uparrow,r'} +\sigma^z_{\downarrow,r} \sigma^z_{\downarrow,r'})\\  &+& J_{t,z} \sigma^z_{\uparrow,d+z}\sigma^z_{\downarrow,d+z},
\label{eq:effHamFull2}
\eeq
where the primed sum, is over all neighboring sites $r$ and $r'$ where $\langle r,r'\rangle \in z$ and $r,r'\neq d$. The $J_z$ terms align the spins at $r$ and $r'$, resulting in one effective spin-$1/2$ for each of the $z$-oriented links on each layer. We will refer to the Pauli matrices for these effective spins as $\bar{\sigma}_{\uparrow}$ for the $z$-oriented links on the top layer and $\bar{\sigma}_{\downarrow}$ for the $z$-oriented links on the bottom layer. These Pauli matrices are defined analogously to those in Eq. \ref{eq:PMDef1}. The $J_{t,z}$ term will also align the two spins at $d+z$ such that $\langle \sigma^z_{\uparrow,d+z} \rangle = \langle \sigma^z_{\downarrow,d+z}\rangle$. Using the same logic as in Sec. \ref{Review}, we can regard the two spins at $d+z$ as a single effective spin-$1/2$. The Pauli matrices for this effective spin are 
\begin{equation}
\begin{split}
&\bar{\sigma}^z_{\updownarrow,d+z} \equiv \sigma^z_{\uparrow,d+z} = \sigma^z_{\downarrow,d+z},\\ &\bar{\sigma}^x_{\updownarrow,d+z} \equiv \sigma^x_{\uparrow,d+z}\sigma^x_{\uparrow,d+z} =\sigma^y_{\uparrow,d+z}\sigma^y_{\uparrow,d+z},\\ &\bar{\sigma}^y_{\updownarrow,d+z} \equiv \sigma^x_{\uparrow,d+z}\sigma^y_{\uparrow,d+z} =\sigma^y_{\uparrow,d+z}\sigma^x_{\uparrow,d+z}.
\end{split}
\label{eq:PMDef2}
\end{equation}
The ground states of Eq. \ref{eq:effHamFull2} thereby consist of a free spin-$1/2$ for every $z$-oriented link on both layers that is not connected to the site $d$, as well as a free spin-$1/2$ shared between both layers at $d+z$. We will now construct a secondary effective Hamiltonian by considering the effects of $J_x,J_y, J_{t,x}$, and $J_{t,y}$ on the effective spin-$1/2$s that make up the ground state of Eq. \ref{eq:effHamFull2}. The secondary effective Hamiltonian is given by 
\beq
 \nonumber H_{\text{eff,}2} = &-&J_{\text{eff}} \sum'_{ p} \left[\bar{\sigma}^z_{\uparrow,i}\bar{\sigma}^y_{\uparrow,j}\bar{\sigma}^z_{\uparrow,k}\bar{\sigma}^y_{\uparrow,l}+ (\bar{\sigma}_{\uparrow} \leftrightarrow \bar{\sigma}_{\uparrow})\right] \\ &\phantom{=}&+ H_{t,d}
\label{eq:kEff2}
\eeq
\beq
\nonumber H_{t,d} = & -&J_{1} (\bar{\sigma}^z_{\uparrow,1}\bar{\sigma}^y_{\uparrow,2}\bar{\sigma}^y_{\uparrow,8}\bar{\sigma}^z_{\downarrow,1}\bar{\sigma}^y_{\downarrow,2}\bar{\sigma}^y_{\downarrow,8})\\ \nonumber &-&J_{2} (\bar{\sigma}^z_{\uparrow,2}\bar{\sigma}^y_{\uparrow,3}\bar{\sigma}^z_{\uparrow,4}\bar{\sigma}^z_{\downarrow,2}\bar{\sigma}^y_{\downarrow,3}\bar{\sigma}^z_{\downarrow,4}\bar{\sigma}^x_{\updownarrow,9})\\\nonumber &-&J_{3} (\bar{\sigma}^z_{\uparrow,8}\bar{\sigma}^y_{\uparrow,7}\bar{\sigma}^z_{\uparrow,6}\bar{\sigma}^z_{\downarrow,8}\bar{\sigma}^y_{\downarrow,7}\bar{\sigma}^z_{\downarrow,6}\bar{\sigma}^x_{\updownarrow,9})\\
&-&J_4 (\bar{\sigma}^z_{\uparrow,5}\bar{\sigma}^y_{\uparrow,4}\bar{\sigma}^z_{\updownarrow,9}\bar{\sigma}^y_{\uparrow,6}+\bar{\sigma}^z_{\downarrow,5}\bar{\sigma}^y_{\downarrow,4}\bar{\sigma}^z_{\updownarrow,9}\bar{\sigma}^y_{\downarrow,6}),
\label{eq:tunEff2}
\eeq
The primed sum indicates a sum over all plaquettes $p$ that do not neighbor the site $d$, and $i$, $j$, $k$, and $l$ label the sites of the effective spins that make up the plaquette $p$ (see Fig. \ref{fig:kEffDia}). These plaquettes are made up entirely of either red or green effective spins in Fig. \ref{fig:ATwistEffDia}. The subscripts in Eq \ref{eq:tunEff2}, correspond to those in Fig. \ref{fig:ATwistEffDia}. All $J_{i}$ are positive, and to leading order, $J_1 \sim \frac{J_x^2J_y^2J_{t,x}J_{t,y}}{J^5_z}$, $J_2 \sim \frac{J_x^3J_y^3J_{t,y}}{J_z^6}$, $J_3 \sim \frac{J_x^3J_y^3J_{t,x}}{J_z^6}$, and $J_4 \sim \frac{J_x^2J_y^2}{J_z^3}$. 

The first two terms in Eq. \ref{eq:kEff2}, describe two independent copies of the Abelian phase of the Kitaev model, i.e., two copies of the toric code. The bulk excitations corresponding to these terms are also just two copies of the toric code anyons, one for each layer. As we shall show, $H_{t,d}$ consists of localized tunneling terms that allow anyons near $d$ to tunnel between the different layers. It is straightforward to verify that all terms in Eq. \ref{eq:kEff2} commute, and that the system will be gapped. The ground state can be found by individually minimizing each of these terms. As with the unperturbed toric code, the ground state of Eq. \ref{eq:kEff2} can be interpreted as a condensate of contractible anyon loops. The main difference is that $H_{t,d}$ allows loops to pass from one layer to the other. 

To explicitly show that $H_{t,d}$ allows bulk anyons from one layer to tunnel to the other layer, let us consider the operator $\bar{\sigma}^z_{\uparrow,2}\bar{\sigma}^z_{\downarrow,2}$, where we have used the subscripts from Fig. \ref{fig:ATwistEffDia}. The operator $\bar{\sigma}^z_{\uparrow,2}\bar{\sigma}^z_{\downarrow,2}$ commutes with $H_{t,d},$ but anti-commutes with the $\bar{\sigma}_{\uparrow}$ and $\bar{\sigma}_{\downarrow}$ plaquette terms located at $\text{P}'$ in Fig. \ref{fig:ATwistEffDia}. This means that $\bar{\sigma}^z_{\uparrow,2}\bar{\sigma}^z_{\downarrow,2}$ will create (or annihilate) a toric code anyon at the plaquette $\text{P}'$ on both the top and bottom layers. Similarly, $\bar{\sigma}^z_{\uparrow,3}\bar{\sigma}^z_{\downarrow,3}$, also commutes with $H_{t,d}$, and anti-commutes with the $\bar{\sigma}_{\uparrow}$ and the $\bar{\sigma}_{\downarrow}$ plaquette terms located at $\text{P}''$ in Fig. \ref{fig:ATwistEffDia}. This means that $\bar{\sigma}^z_{\uparrow,3}\bar{\sigma}^z_{\downarrow,3}$ will create (or annihilate) a toric code anyon at the plaquette $\text{P}''$ on both the top and bottom layers. Since all toric code anyons are their own anti-particles, these processes allow toric code anyons to tunnel from one layer to the other. 

These tunneling processes can either send $e$($m$) anyons on one layer to $e$($m$) anyons on the other layer, or send $e$($m$) anyons on one layer to $m$($e$) anyons on the other layer. To show this, we first recall that the $e$ and $m$ toric code anyons are localized on alternating rows of hexagonal plaquettes of the honeycomb lattice. Without loss of generality, let us consider the case where the $\bar{\sigma}_{\uparrow}$ plaquette term at $\text{P}'$ corresponds to an $e$ anyon, and the $\bar{\sigma}_{\uparrow}$ plaquette at term $\text{P}''$ corresponds to an $m$ anyon. In this case, it is possible to have the $\bar{\sigma}_{\downarrow}$ plaquette term at $\text{P}'$ correspond to either an $e$ anyon or $m$ anyon (and the $\bar{\sigma}_{\downarrow}$ plaquette term at $\text{P}''$ correspond to an $m$ anyon or $e$ anyon respectively). In the former case, $\bar{\sigma}^z_{\uparrow,2}\bar{\sigma}^z_{\downarrow,2}$ will tunnel an $e$ anyon on the top layer into an $e$ anyon on the bottom layer, and $\bar{\sigma}^z_{\uparrow,3}\bar{\sigma}^z_{\downarrow,3}$ will tunnel an $m$ anyon on the top layer into an $m$ anyon on the bottom layer. In the latter case $\bar{\sigma}^z_{\uparrow,2}\bar{\sigma}^z_{\downarrow,2}$ will tunnel an $e$ anyon on the top layer into an $m$ anyon on the bottom layer, and $\bar{\sigma}^z_{\uparrow,3}\bar{\sigma}^z_{\downarrow,3}$ will tunnel an $m$ anyon on the top layer into an $e$ anyon on the bottom layer. These different types of tunneling processes will be of interest when studying full twisted bilayers with coupled spins on all coincident sites. 

Regardless of the details of the tunneling processes, we can confirm that $H_{t,d}$ allows the non-trivial toric code anyons of one layer to tunnel into the other layer. Because of this, $H_{t,d}$ can be interpreted as a pair of genon defects\cite{barkeshli2013}. In a bilayer system, a genon defect is the end point of a branch cut that connects the two layers. When an anyon from one layer passes through this branch cut, it becomes an anyon on the other layer. Since, $H_{t,d}$ allows anyons to tunnel from one layer to the other, it constitutes a single isolated segment of such a branch cut, or equivalently, a pair of genon defects (one at each end of the branch cut). Because of this, we will refer to the defect in Eq. \ref{eq:tunEff2} as a "bi-genon" defect. 

\subsection{Defect Lattice}
\label{SSDefectLat}
We will now turn our attention to the commensurate twisted Kitaev bilayers, where spins on each coincident site are coupled. To avoid any effects caused by the boundaries or non-trivial topology of the lattice, we will consider the case where each lattice layer is an infinite plane. As noted before, we will assume that the interlayer interactions are significantly short-ranged, so that only the spins that are located on coincident sites are coupled. When the bilayer is twisted by a commensurate twist angle $\theta(m,r)$, the interlayer interaction is given by:
\beq
H_{\text{inter}} =K \sum_{r \in C} \vec{\sigma}_{\uparrow,r}\cdot \vec{\sigma}_{\downarrow,r},
\label{eq:sLinkLat}
\eeq
where $C$ is the set of coincident sites of the twisted bilayer system (black sites in Fig. \ref{fig:TwistLat}). The sites $r \in C$ form a super-lattice with primitive vectors given by Eq. \ref{eq:PV1} or \ref{eq:PV2}. Again, we will consider the limit $K \gg J_z$. Following the same logic as before, we find that the effective Hamiltonian for the twisted bilayer system is
\beq
\nonumber H_{\text{twist}} = &-&J_{\text{eff}} \sum'_{ p}[ \bar{\sigma}^z_{\uparrow,i}\bar{\sigma}^y_{\uparrow,j}\bar{\sigma}^z_{\uparrow,k}\bar{\sigma}^y_{\uparrow,l} + (\bar{\sigma}_{\uparrow} \leftrightarrow \bar{\sigma}_{\uparrow})]\\ &+& \sum_{r \in C} H_{t,r}.
\label{eq:EffTwistHam}
\eeq
Here the primed sum indicates a sum over all plaquettes $p$ that do not neighbor any of the spins that are coupled by Eq. \ref{eq:sLinkLat}. As before, the first and second terms are the bulk toric code couplings. $H_{t,r}$ is a tunneling interaction that is localized at the coincident site $r\in C$. The tunneling term $H_{t,r}$ is defined analogously to $H_{t,d}$ in Eq. \ref{eq:tunEff2}. 

From our previous analysis, we know that $H_{t,r}$ constitutes a localized bi-genon defect that allows anyons to pass from one layer to the other. All terms in Eq. \ref{eq:EffTwistHam} commute with each other, and the ground state can be found by individually minimizing each term. As before, we can interpret the ground state of Eq. \ref{eq:EffTwistHam} as a condensate of contractible closed loops of the bulk toric code anyons. Due to the bi-genon defects, these loops may pass from one layer to the other.  Since there is a bi-genon located at each of the coincident sites ($r\in C$) of the twisted bilayer, Eq. \ref{eq:EffTwistHam} can be interpreted as two copies of toric code, which are coupled by a super-lattice of bi-genons. The bi-genons form a triangular lattice for SE-odd twist angles, and a hexagonal lattice for SE-even twist angles. As noted previously, there are two types of bi-genon defects in the bilayer system. Those that take $e$ anyons to $e$ anyons and $m$ anyons to $m$ anyons (which will be referred to as "$ee$" defects) and those that take $e$ anyons to $m$ anyons and $m$ anyons to $e$ anyons (which will be referred to as "$em$" defects). For SE-odd twist angles, the $ee$ and $em$ each form one of the two inter-penetrating rectangular lattices that make up the triangular lattice. For SE-odd twist angles, the $ee$ and $em$ each form one of the two inter-penetrating triangular lattices that make up the hexagonal lattice. This can be confirmed from Eqs. \ref{eq:PV1} and \ref{eq:PV2}, and by noting that since the $e$ and $m$ anyons are defined on alternating rows of plaquettes of the honeycomb lattice, translating a single lattice layer by $\textbf{a}_1$ or $\textbf{a}_2$ (see Fig. \ref{fig:kLinks}) exchanges the $e$ and $m$ anyons on that layer.

\begin{figure}
\centering
\includegraphics[width = \linewidth*3/4]{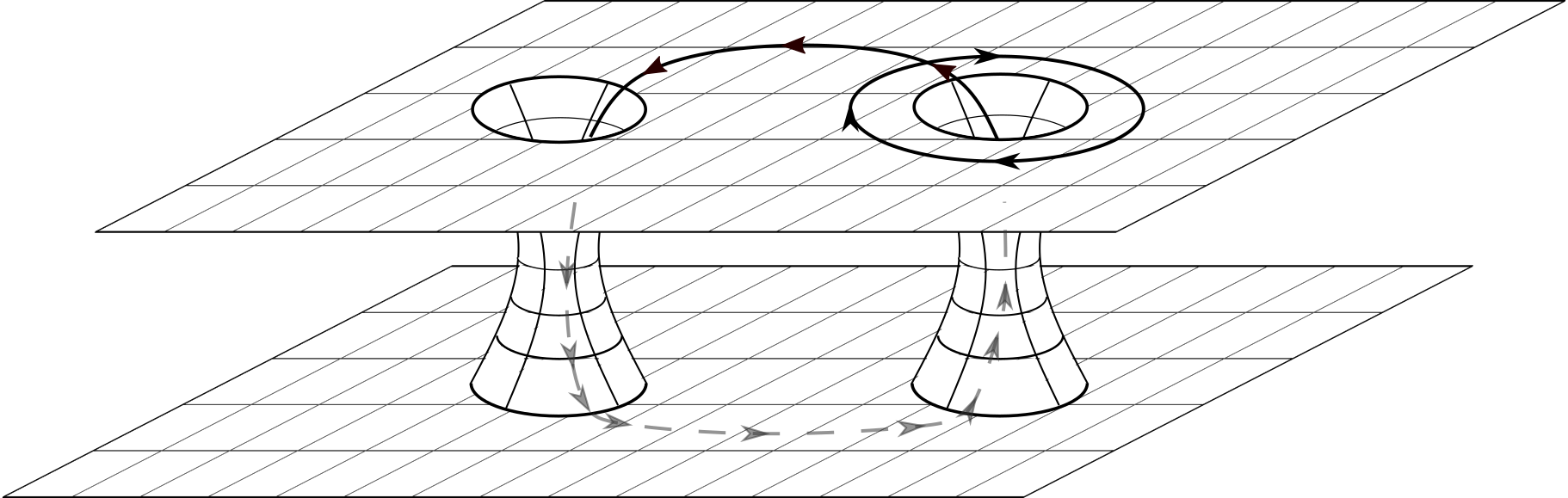}
\caption{A pair of lattice defects that allow anyons from one layer to pass through to the other layer. Here we indicate the presence of a new pair non-contractible paths caused by the defects.}
\label{fig:DefectDia}
\end{figure}

A key feature of this bi-genon super-lattice is that each bi-genon allows for two new types of non-contractible anyon loops. First, there are the non-contractible anyon loops that remain on a given layer and encircle a bi-genon. Second, there are the non-contractible anyon loops that pass from one layer to the other through a bi-genon, and then return to the original layer through a different bi-genon. These non-contractible anyon loops are shown in Fig. \ref{fig:DefectDia}. The operators that create these non-contractible anyon loops commute with Eq. \ref{eq:EffTwistHam}, and make up the non-trivial low energy degrees of freedom of the theory. In general, these non-contractible anyon loops do not commute with each other, and they form a non-trivial algebra. Because of this, it can be shown that each bi-genon increases the ground state degeneracy of the system by a factor of $4$\cite{barkeshli2013}. For a system of area $A$ and a density of bi-genons $\rho$, the ground state degeneracy due to the bi-genons is $4^{\rho A}$. 

\section{The Moir\'e Ising Model}
\label{MIsingModel}
We now turn our attention to the low-energy dynamics associated with the defect lattice from Sec. \ref{SSDefectLat}. As noted before, non-contractible anyon loops make up the zero energy, non-trivial, degrees of freedom of the coupled bilayers in Eq. \ref{eq:EffTwistHam}. We are interested in finding the effective Hamiltonian for these anyon loops. To imbue these loops with dynamics, we will include the following perturbation to the bilayer system,
\beq
H_{\text{mag}} = - \sum_i [h\bar{\sigma}^x_{\uparrow, i} + h\bar{\sigma}^y_{\uparrow, i}+ h \bar{\sigma}^z_{\uparrow, i}] + (\bar{\sigma}_{\uparrow} \leftrightarrow \bar{\sigma}_{\downarrow}),
\label{eq:HDyn}
\eeq
where $|h|$ is significantly smaller than all other energy scales in Eq. \ref{eq:EffTwistHam}. In the presence of Eq. \ref{eq:HDyn}, anyon loops can be dynamically created by virtual processes where pairs of anyons are created and annihilated, forming closed anyon loops. The amplitude to create a given anyon loop will be proportional to $(h/E_{\text{gap}})^l$, where, $E_{\text{gap}}$ is the energy gap of the anyon, and $l$ is the length of the loop in units of the lattice constant. At this point, we would like to note a difference between the $e$ and $m$ anyons and the $\psi$ anyons in toric code (Eq. \ref{eq:kEff} and \ref{eq:TorHam}). In the lattice model, both the $e$ and $m$ anyons have a gap of $2 J_{\text{eff}}$. Since the $\psi$ anyon is a fusion of an $e$ anyon and an $m$ anyon, it has a gap of $4J_{\text{eff}}$. The amplitude to create a loop of $\psi$ anyons is thereby reduced by a factor of $1/2^l$, compared to a loop of $e$ or $m$ anyons of the same length. Due to this extra suppression we will ignore processes that create, $\psi$ anyon loops. 

The leading contributions to the effective Hamiltonian will therefore come from short non-contractable loops of $e$ and $m$ anyons. The four shortest types of non-contractible loops are: loops that encircle a single $ee$ defect, loops that pass through a neighboring pair of $ee$ defects, loops that encircle a single $em$ defect, and loops that pass through a neighboring pair of $em$ defects. Since all anyon loops of the first type and second type commute with all anyon loops of the third type and fourth type, the $ee$ and $em$ defects decouple at leading order. Because of this, we will only consider a single type of defect for our analysis. Without loss of generality we will chose to focus on the $ee$ defects. The analysis for the $em$ defects is identical.

\begin{figure}
\centering
\includegraphics[width = 0.8\columnwidth]{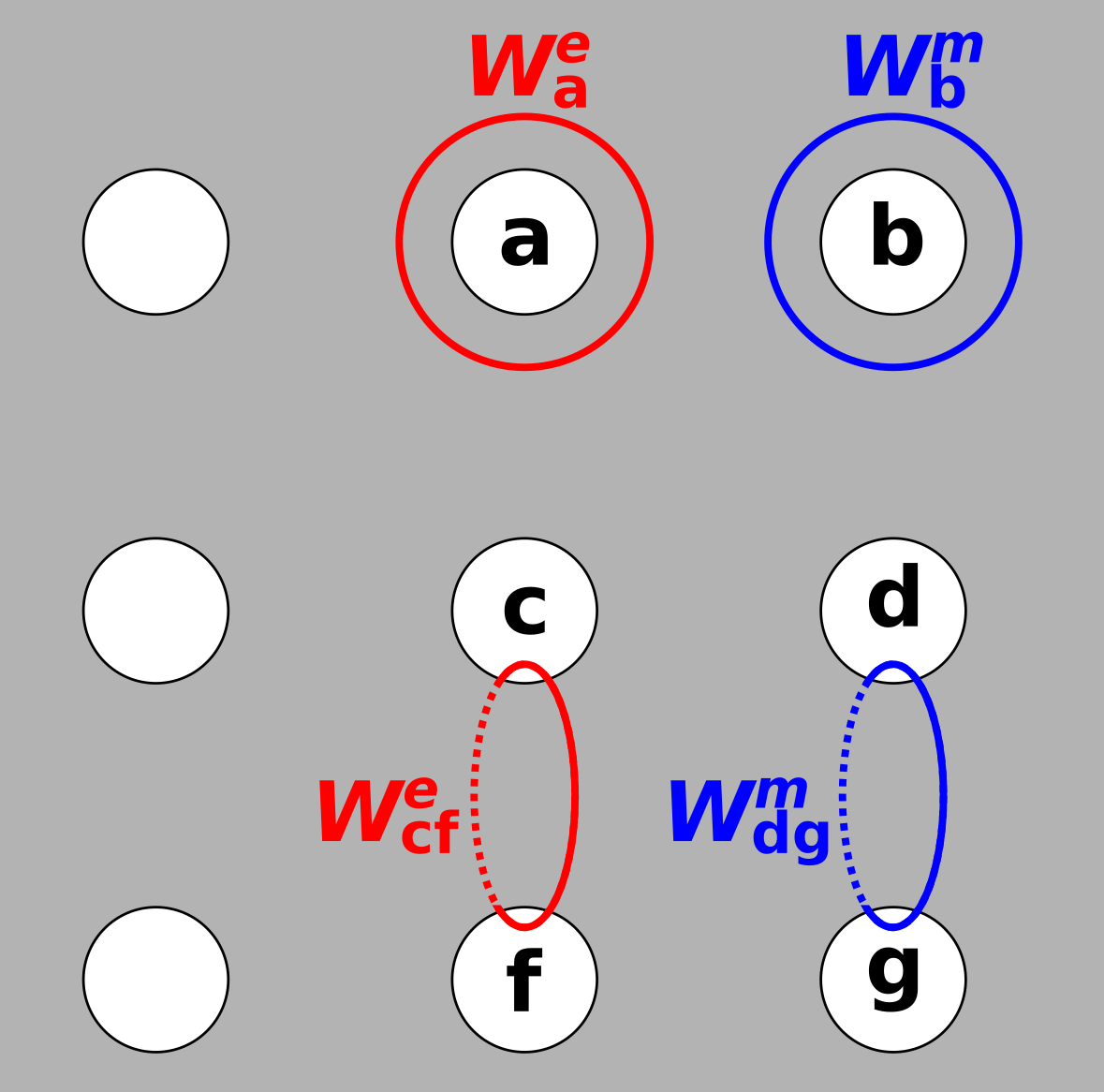}
\caption{A square lattice of $ee$ defects (white) that form a rectangular sublattice of an underlying triangular lattice.  The nontrivial $e$ (red) and $m$ (blue) anyon loops are shown. These loops can either circle a defect ($W^e_{a}$ and $W^m_{b}$) or pass through a neighboring pair of defects ($W^e_{cf}$ and $W^m_{dg}$).}
\label{fig:LatDefect3}
\end{figure}

In terms of the creation and annihilation operators for the anyon loops, the effective Hamiltonian for the $ee$ defect lattice is given by
\beq
\nonumber H_{\text{defect}} =& -&\sum_{i} (u^e_{i} W^e_{i} + u^m_{i} W^m_{i})\\ &-&\sum_{\langle i j\rangle} (t^{e}_{i j} W^{e}_{i j} + t^{m}_{i j} W^{m}_{i j}),
\label{eq:loopHam0}
\eeq
where $i$ labels the position of an $ee$ defect in the lattice, and $ \langle i j\rangle$ indicates a neighboring pair of defects. The operator $W^{e(m)}_{i}$ creates a loop of $e(m)$ anyons that surrounds the $i^{th}$ defect. The operator $W^{e(m)}_{i j}$ creates a loop of $e(m)$ anyons that passes from one layer to the other through the defect at site $i$ and returns to the original layer through the defect at site $j$. The coupling $u^{e(m)}_{i}$ is the amplitude to create an $e(m)$ loop around the $i^{\text{th}}$ defect, and $t^{e(m)}_{ij}$ is the amplitude to create an $e(m)$ loop that passes through the $i^{\text{th}}$ and $j^{\text{th}}$ defects. The defect Hamiltonian is summarized in Fig.  \ref{fig:LatDefect3}. The values of $u^{e(m)}_{i}$ and $t^{e(m)}_{ij}$ depend the twist angle between the bilayers and on positions of the $i^{\text{th}}$ and $j^{\text{th}}$ defects. The dependence of the coupling constants on position is due to two features of the Kitaev bilayer system. First, since we are in the Abelian phase of the Kitaev model, we have explicitly broken rotational symmetry ($J_z \gg J_x,J_y$). As a result, the Ising couplings can be anisotropic. Second, as noted before, the $e$ and $m$ anyons of the Kitaev model are defined on alternating rows of hexagonal plaquettes. This extra periodic structure can lead to both the Ising couplings and the transverse magnetic field terms being periodic with period-$2$. We see both of these effects when considering explicit examples in Appendix \ref{App:MIM}.

Using the braiding relationships between the $e$ and $m$ anyons, we find that the loop operators obey the commutation relationships $\{W^e_i, W^m_{ij} \} = \{W^e_j, W^m_{ij} \} = 0$, and $\{W^m_i, W^e_{ij} \} = \{W^m_j, W^e_{ij} \} = 0$. All other loop operators commute. Additionally, since the $e$ and $m$ anyons are both their own anti-particle, the anyon loops satisfy $(W^e_i)^2 = (W^m_i)^2 = (W^e_{ij})^2= (W^m_{ij})^2 = 1$. This algebra is equivalent to having two spin-$1/2$ degrees of freedom at each defect site. To show this, we will define two sets of Pauli matrices at the defect site $i$: $\tau_{1,i}$ and $\tau_{2,i}$. The anyon loop algebra can then be satisfied by the following identifications: $W^e_{i} = \tau^x_{1,i}$, $W^m_{i} = \tau^x_{2,i}$, $W^m_{ij} = \tau^z_{1,i}\tau^z_{1,j}$ and $W^e_{ij} = \tau^z_{2,i}\tau^z_{2,j}$. The defect lattice can then be treated as a lattice system with two spins-$1/2$s per site. In terms of the new spin degrees of freedom, Eq. \ref{eq:loopHam0} is
\beq
\nonumber H_{\text{defect}} &=& -\sum_{i} (u^e_{i} \tau^x_{1,i} +u^m_{i} \tau^x_{2,i})\\ &-&\sum_{\langle i j\rangle} (t^{e}_{i j} \tau^z_{2,i}\tau^z_{2,j} + t^{m}_{i j} \tau^z_{1,i}\tau^z_{1,j} ),
\label{eq:loopHamI0}
\eeq
which includes all leading order contributions to the dynamics of the $ee$ defect lattice. Eq. \ref{eq:loopHamI0} can be readily identified as two decoupled quantum Ising models. There will be another two Ising models arising from the $em$ defects, leading to a total of four decoupled Ising models. As noted before, the Ising couplings ($t^{e(m)}_{i j}$) and transverse magnetic field terms ($u^{e(m)}_{i}$) in Eq. \ref{eq:loopHamI0} both depend on position.  

In Appendix \ref{App:MIM} we explicitly analyze Eq. \ref{eq:loopHamI0} when the twist angle is given by $(m,r) = (1,1)$ and when the twist angle is given by $(m,r) = (2,1)$. We find that for the $(m,r) = (1,1)$ twist angle both the $\tau_{1}$ and $\tau_{2}$ spins are in a ferromangetic phase, where $\langle \tau^z_{1,i} \rangle = \pm 1$, and $\langle \tau^z_{2,i} \rangle = \pm 1$. This leads to a total of four groundstates for the $ee$ defect sector. There are also another four ground states arising from the $em$ defects for a total of $4 \times 4 = 16$ ground states for the entire defect lattice system. For $(m,r) = (2,1)$ twist angle, both the $\tau_{1}$ and $\tau_{2}$ spins are in a paramagnetic phase where $\langle \tau^x_{1,i} \rangle =\langle \tau^x_{2,i} \rangle = 1$, and the ground state of the system of $ee$ defects is unique. The ground state is also unique for the system of $em$ defects, and so ground state of the entire defect lattice system is unique for the $(m,r) = (2,1)$ twist angle. In this analysis, we have used the fact that to leading order, the $ee$ and $em$ defect lattices decouple. We consider subleading terms that couple the $ee$ and $em$ defect sectors in Appendix \ref{App:Sub}. In particular we find that for the $(m,r) = (1,1)$ twist angle, subleading couplings between the $ee$ and $em$ defects reduce the number of ground states from $16$ to $8$. 

Since the effective Ising spins correspond to non-contractible anyon loops, the magnetization order parameters that characterize the ground state of the effective Ising models in Eq. \ref{eq:loopHamI0} are extended spin operators in terms of the original Kitaev degrees of freedom. These extended spin operators are the product of a finite number of spin operators ($\sigma_{\uparrow,r}$ and $\sigma_{\downarrow,r}$ in Eq. \ref{eq:kModel2}) that act on different sites of the bilayer system. Because of this, one would have to simultaneously measure multiple spins in the bilayer system in order to measure the effective Ising order parameter, which may be challenging experimentally, but could likely be carried out as a numerical diagnostic.

This example demonstrates the non-trivial effect of the twist angle on the the emergent spin  model. In general, the defects will be further apart from each other for smaller twist angles. This will suppresses the amplitude to create a loop that connects neighboring defects ($t^e_{ij}$ and $t^m_{ij}$ in Eq. \ref{eq:loopHam0}). On the other hand, the distance around a given defect is independent of the twist angle, and the amplitude to create a loop that encirlces a given defect ($u^e_{i}$ and $u^m_{i}$ in Eq. \ref{eq:loopHam0}) will not be suppressed as the twist angle is decreased. Based on this, we expect that at small twist angles the $u^e_{i}$ and $u^m_{i}$ terms will dominate in \ref{eq:loopHamI0}, and the emergent Ising models will be in a trivial paramagnetic phase.

\section{Conclusion and Outlook}
\label{Conclusion}
We have shown that coupling twisted Kitaev model bilayers can lead to new emergent physics at low energies. At commensurate twist angles, coupling coincident spins in the two layers generates a lattice of bi-genon defects. These bi-genons allow anyons to pass from one layer to the other, and lead to a ground state degeneracy that is exponential with system size. This ground state degeneracy can be split by applying a weak magnetic field to the system. The resulting low energy physics is well described by four 2D quantum Ising models. Furthermore, we have shown that at large twist angles, the emergent Ising models are in a ferromagnetic phase, and at small twist angles, the Ising models are in a paramagnetic phase. In this analysis, we have only explicitly considered SE-odd twist angles. The results for SE-even twist angles are analogous.

It may be of interest to extend this procedure to the non-Abelian phase of the Kitaev model. In doing so we hope to be able to describe the low energy physics that may be found in twisted versions of proposed Kitaev materials such as $\alpha$-RuCl$_3$. However, for $\alpha$-RuCl$_3$ the interlayer distance is large, and the strong coupling approach we used here may not apply. Our methodology may also prove useful in analyzing heterostructures of spin liquids where the lattices of the two layers are mismatched\cite{singh1993}. In lattice mismatched heterostructures, a moir\'e pattern and superlattice can also form, and our results would be applicable. The approach we have used here may provide insight into the low energy excitations that have been found in the random Kitaev magnet CU$_2$IrO$_3$\cite{choi2019}. It may also be of interest to study other topologically ordered systems where a super lattice of twist defects has been added. To our knowledge, the dynamics associated with such a system have not yet been explored. 

\section*{Acknowledgments}
We thank H. Goldman for helpful discussions. JMM is supported by the National Science
Foundation Graduate Research Fellowship Program under Grant No. DGE 1746047. TLH thanks the US National Science Foundation under
grant DMR 1351895-CAR
for support. TLH also thanks the
National Science Foundation under Grant No.NSF PHY1748958(KITP) for partial support at the end stage of
this work during the Topological Quantum Matter program.

\begin{appendix}

\section{The Moir\'e Ising Model at $(m,r) = (1,1)$ and $(m,r) = (2,1)$ Twist Angles}\label{App:MIM}

We will now explicitly consider two cases of the Moir\'e Ising model found Sec. \ref{MIsingModel}. First, when the twist angle is given by $(m,r) = (1,1)$ ($\theta(1,1) \approx 21.79^o$), and second, when the twist angle is given by $(m,r) = (2,1)$ ($\theta(2,1) = 13.17^o$). 

\begin{figure}
\centering
\includegraphics[width = \linewidth*3/4]{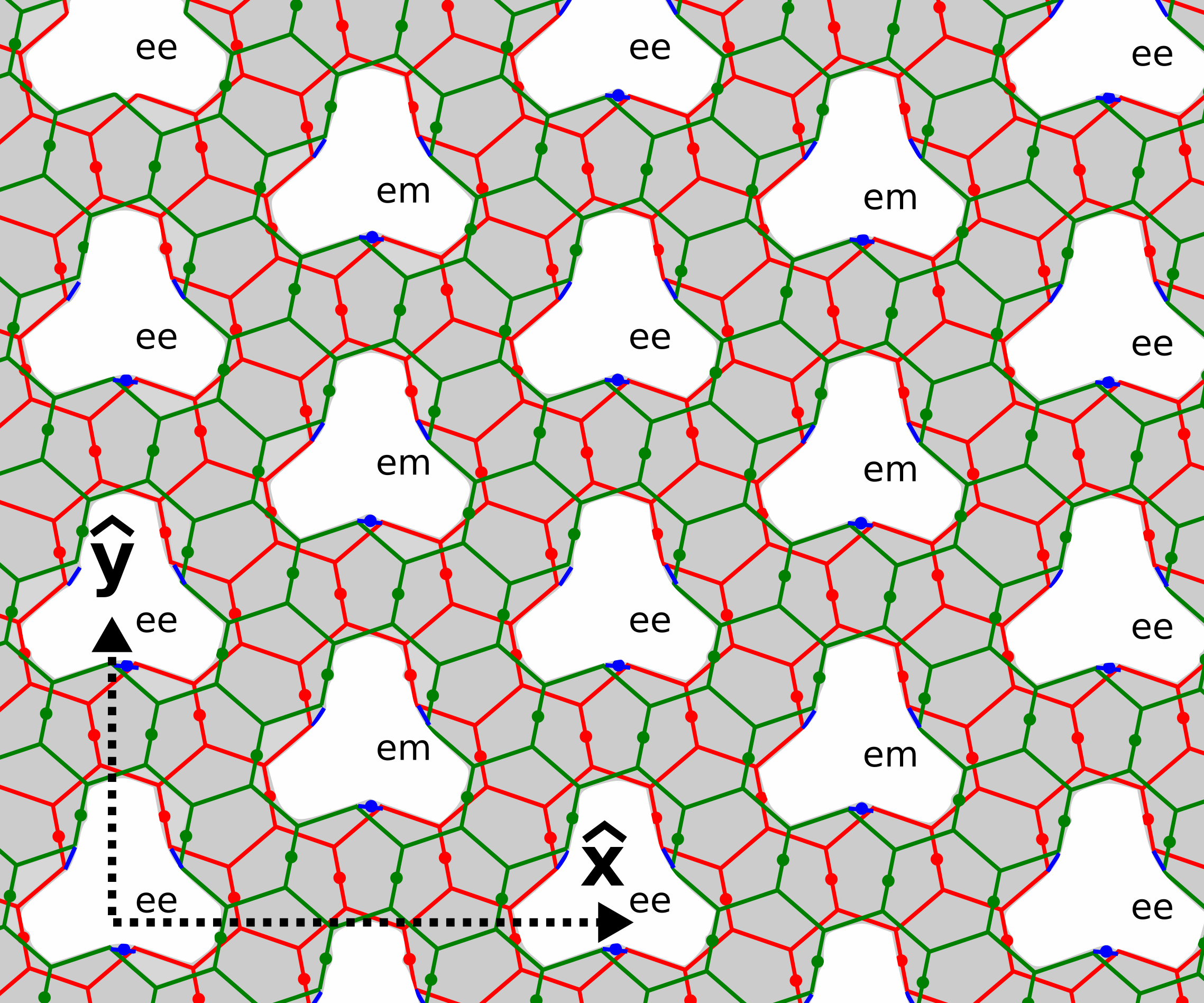}
\caption{The defect lattice for the $(m,r) = (1,1)$ twist angle. There are two types of defects, $ee$ and $em$, each of which form a rectangular lattice with primitive vectors $\hat{x}$ and $\hat{y}$.}
\label{fig:TwistLat2}
\end{figure}

The $(m,r) = (1,1)$ twist angle is SE-odd and the aligned spins form a triangular lattice. As stated before, there are two types of bi-genon defects for the $(m,r) = (1,1)$ twist pattern, $ee$ and $em$ defects. The $ee$ and $em$ defects each make up a rectangular sub-lattice of the triangular super-lattice (see Fig. \ref{fig:TwistLat2}). Since the $ee$ and $em$ defects do not couple to each other at leading order, we will only consider the rectangular lattice of $ee$ defects. We will label the primitive vectors of the $ee$ defect lattice as $\hat{x}$ and $\hat{y}$ as shown in Fig. \ref{fig:TwistLat2}. Using the same logic as before, the Hamiltonian for the $ee$ defects is given by Eq. \ref{eq:loopHamI0} where $i$ and $j$ label the positions of the $ee$ defects in the $(m,r) = (1,1)$ defect lattice. For our purposes, it will be useful to rewrite Eq. \ref{eq:loopHamI0} as 
\beq
\nonumber H_{\text{defect}} &=& -\sum_{i} ( u^e_{i} \tau^x_{1,i} + u^m_{i} \tau^x_{2,i})\\\nonumber &-&\sum_{\langle i j\rangle \in x} (t^{e,\perp}_{i j} \tau^z_{2,i}\tau^z_{2,j} + t^{m,\perp}_{i j} \tau^z_{1,i}\tau^z_{1,j} )\\ &-&\sum_{\langle i j\rangle \in y} (t^{e,\parallel}_{i j} \tau^z_{2,i}\tau^z_{2,j} + t^{m,\parallel}_{i j} \tau^z_{1,i}\tau^z_{1,j} ),
\label{eq:loopHamI1}
\eeq
where $\langle i j\rangle \in x$ indicates nearest neighbor pair of $ee$ defects in the $\hat{x}$-direction, and $\langle i j\rangle \in y$ indicates nearest neighbor pair of defects in the $\hat{y}$-direction. All coupling constants in Eq. \ref{eq:loopHamI0} are positive, and their approximate magnitudes are given in Table \ref{tab:Coeffs}. We note that these couplings are anisotropic, and additionally are periodic with period-$2$. 

From Table \ref{tab:Coeffs}, it is clear that for $(m,r) = (1,1)$, the $t^{e,\perp}_{ij}$ and $t^{m,\perp}_{ij}$ couplings are smaller than all other terms in the effective Hamiltonian, and can be treated as perturbations. In the absence of the intercolumn couplings, each column of Eq. \ref{eq:loopHamI1} consists of two decoupled Ising chains oriented along the $\hat{y}$-direction. The Hamiltonian for a single column, which we will label as $I$, is given by
\begin{equation}
\begin{split}
H_{I} = -\sum_{i\in I} \big[&u^e_{i} \tau^x_{1,i}+ t^{m,\parallel}_{i, i+\hat{y}} \tau^z_{1,i}\tau^z_{1,i+\hat{y}}  \\ & u^m_{i} \tau^x_{2,i} + t^{e,\parallel}_{i, i+\hat{y}} \tau^z_{2,i}\tau^z_{2,i+\hat{y}}\big],
\end{split}
\label{eq:loopHamI2}
\end{equation}
where the sum is only over the spins in the column $I$. Since the magnitudes of $u^{e/m}_i$, and $t^{e/m,\parallel}_{i j}$, all alternate with period $2$, Eq. \ref{eq:loopHamI2} consists of a pair of period-$2$ Ising chains. Period- $p$ Ising chains have been previously studied by Derzhko et al.\cite{derzhko2002, derzhko2004}, and their phase diagrams are known. Using their results, we find that for Eq. \ref{eq:loopHamI2}, the $\tau_1$ spins of the column $I$ are in a trivial paramagnetic phase if $\prod_{i \in {I}} t^{m,\parallel}_{i,i+\hat{y}}<\prod_{i \in {I} } u^{e}_i$, and a symmetry broken ferromagnetic phase if $\prod_{i \in {I}} t^{m,\parallel}_{i,i+\hat{y}}>\prod_{i \in {I} } u^{e}_i$. In the paramagnetic phase, $\langle \tau^x_{1,i} \rangle_{I} = 1$, ($\langle...\rangle_{I} $ indicates an average taken over the spins in the column $I$) and in the ferromangetic phase $\langle \tau^z_{1,i} \rangle_{I}  = \pm1$.  Similarly, the $\tau_2$ spins of the $I$ column are in a paramagnetic phase if $\prod_{i \in {I}} t^{e,\parallel}_{i,i+\hat{y}}<\prod_{i \in {I} } u^{m}_i$, and a ferromagnetic phase if $\prod_{i \in {I}} t^{e,\parallel}_{i,i+\hat{y}}>\prod_{i \in {I} } u^{m}_i$. Using Table \ref{tab:Coeffs}, we find that both the $\tau_1$ and $\tau_2$ spins in column $I$ are in the ferromagnetic phase for the $(m,r) = (1,1)$ twist angle (recall that $J_{1,2} \ll J_{eff}$).

We will now consider the effects of the sub-leading $t^{e,\perp}_{ij}$ and $t^{m,\perp}_{ij}$ terms in Eq. \ref{eq:loopHamI1}. As we have shown, in the absence of $t^{e,\perp}_{ij}$ and $ t^{m,\perp}_{ij}$, the $\tau_1$ spins of each column are ferromagnetically aligned, as are the $\tau_2$ spins. It is then straightforward to see that the energy associated with $t^{m,\perp}_{ij}$ is minimized when the $\tau_1$ spins on neighboring columns are ferromangetically aligned as well. Similarly, the energy associated with $t^{e,\perp}_{ij}$ is minimized when the $\tau_2$ spins on neighboring columns are also aligned. Because of this, in the ground state of Eq. \ref{eq:loopHamI1}, all $\tau_1$ spins will be ferromangetically aligned ($\langle \tau^z_{1,i}\rangle = \pm 1$) and all $\tau_2$ spins will be ferromangetically aligned ($\langle \tau^z_{2,i}\rangle = \pm 1$). This will result in a total of $4$ ground states for the lattice of $ee$ defects. The analysis for the $em$ defects is identical, and the system of $em$ defects also has $4$ ground states for the same reasons. There will thereby be a total of $4\times 4 = 16$ ground states for the system.

\begin{table}[]
\begin{tabular}{|rl|}
\hline
                                   Coupling &constants for the $(m,r)=(1,1) $ twist angle \\ \hline
$u^e_i$                             & $ \sim \frac{h^4}{J_{\text{eff}}^2 J_2}$ for $i$ even  \\
                                    & $\sim \frac{h^6}{J_{\text{eff}}^5}$ for $i $ odd       \\ \hline
$u^m_i$                             & $\sim \frac{h^6}{J_{\text{eff}}^5}$ for $i $ even      \\ 
                                    & $ \sim \frac{h^4}{J_{\text{eff}}^2 J_2 }$ for $i$ odd  \\ \hline
$t^{e,\parallel}_{ij}$ & $\sim \frac{h^3}{J_1J_2}$ for $i$ even         \\ 
                                    & $\sim \frac{h^7}{J_{\text{eff}}^5J_1}$ for $i$ odd  \\\hline 
$t^{m,\parallel}_{ij}$ & $\sim \frac{h^7}{J_{\text{eff}}^5J_1}$ for $i$ even\\ 
                                    & $\sim \frac{h^3}{J_1J_2}$ for $i$ odd      \\\hline
$t^{m,\perp}_{ij}$ & $\sim \frac{h^9}{J_{\text{eff}}^5J_1J_2J_3}$  for all $i$ \\ \hline $t^{e,\perp}_{ij}$ & $\sim \frac{h^9}{J_{\text{eff}}^5J_1J_2J_3}$  for all $i$ \\ \hline
\end{tabular}

\caption{The perturbative values of the coefficients used in Eq. \ref{eq:loopHam0} for the  $(m,r) = (1,1)$ twist angle. The lattice site $i = (x,y)$ is even when $x +y = 0 \text{ mod}(2)$ and odd when $x +y = 1 \text{ mod}(2)$.}
\label{tab:Coeffs}
\end{table}

\begin{figure}
\centering
\includegraphics[width = \linewidth*3/4]{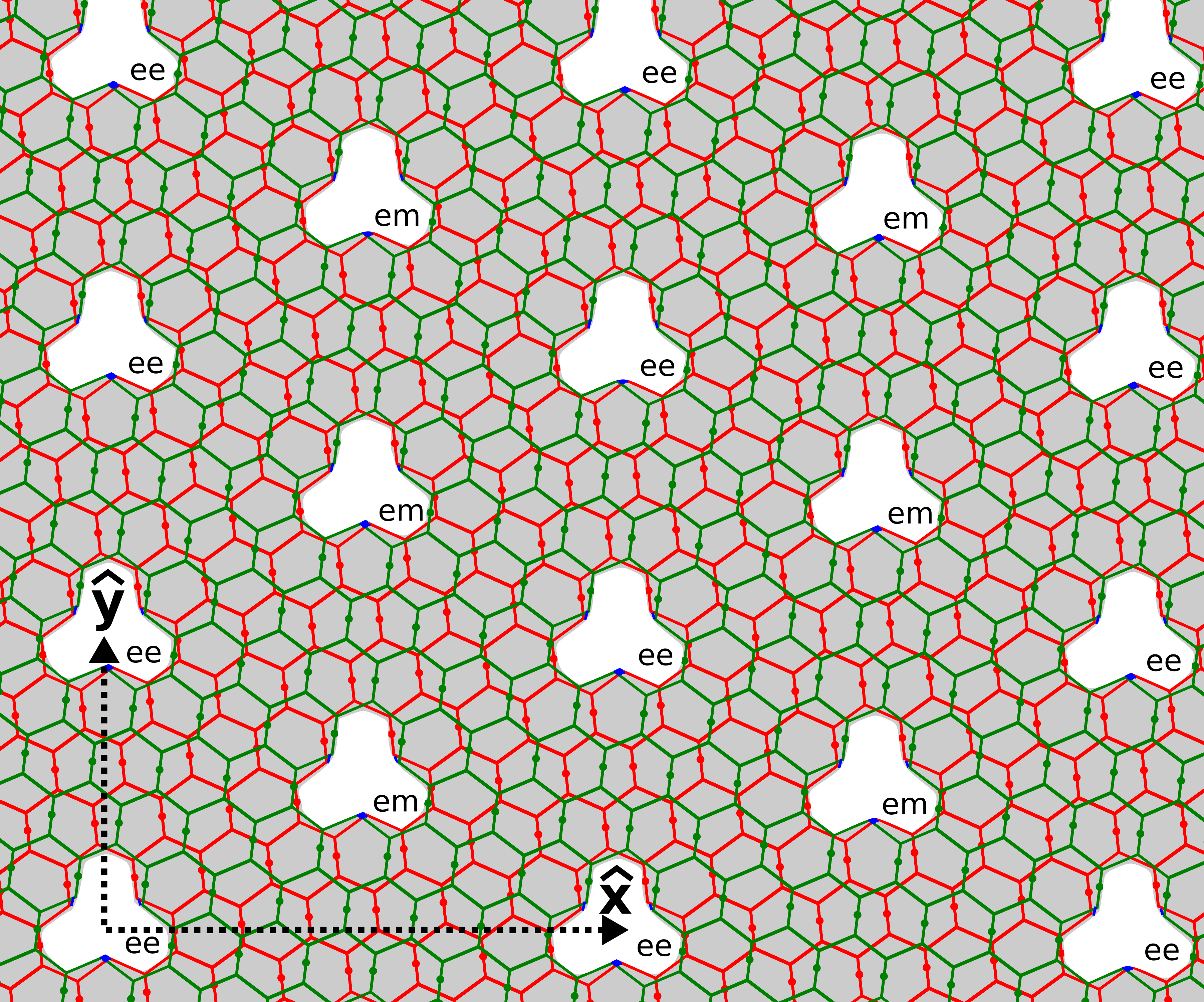}
\caption{The defect lattice for the $(m,r) = (2,1)$ twist angle. There are two types of defects,$ee$ and $em$, each of which form a rectangular lattice with primitive vectors $\hat{x}$ and $\hat{y}$.}
\label{fig:TwistLat4}
\end{figure}

We will now turn our attention to the $(m,r) = (2,1)$ twist angle. Similar to the $(m,r) = (1,1)$ twist angle, the $(m,r) = (2,1)$ twist angle forms SE-odd super-lattice, which is shown in Fig \ref{fig:TwistLat4}. We will analyze this system in the same way that we analyzed the $(m,r) = (1,1)$ system. As with the  $(m,r) = (1,1)$ twist angle, the $ee$ and $em$ defects of the $(m,r) = (2,1)$ twist angle each make up a rectangular sub-lattice. The Hamiltonian for the rectangular lattice of $ee$ defects takes the same form as Eq. \ref{eq:loopHamI1}, where $i$ and $j$ now label the positions of the $ee$ defects in the $(m,r) = (2,1)$ defect lattice. For $(m,r) = (2,1)$ all coupling constants are positive, and the approximate magnitude of the coupling constants at this twist angle are given in Table \ref{tab:Coeffs2}. These couplings are also anisotropic and are periodic with period-$2$. 

As before, the couplings $t^{e,\perp}_{ij}$ and $t^{m,\perp}_{ij}$ for the $(m,r) = (2,1)$ twist angle are subleading, and we can treat them as perturbations. When $t^{e,\perp}_{ij}=t^{m,\perp}_{ij}=0$, we can again consider the Hamiltonian for a single column of spins (see Eq. \ref{eq:loopHamI2}). For the $(m,r) = (2,1)$ twist angle, we find that for each column $I$, $\prod_{i \in I} t^{m,\parallel}_{i,i+\hat{y}}<\prod_{i \in I} u^{e}_i$ and $\prod_{i \in I} t^{e,\parallel}_{i,i+\hat{y}}<\prod_n u^{m}_i$. Following the same logic as before, the $\tau_1$ and $\tau_2$ spins of each column will be in the paramagnetic phase. Since the spins of each column are in the paramagnetic phase, $ \langle \tau^x_{1,i}\rangle = \langle \tau^x_{2,i}\rangle = 1$ for \textit{all} spins in the system, and the ground state is unique. It is clear that the weak ferromagnetic couplings, $t^{e,\perp}_{ij}$ and $t^{m,\perp}_{ij}$ are irrelevant here. The same analysis also applies to the $em$ defects of the $(m,r)=(2,1)$ twist angle. So the ground state of the entire system is unique.

\begin{table}[]
\begin{tabular}{|rl|}
\hline
                                   Coupling &constants for the $(m,r)=(2,1) $ twist angle  \\ \hline
$u^e_i$                             & $ \sim \frac{h^4}{J_{\text{eff}}^2 J_2}$ for $i$ even  \\
                                    & $\sim \frac{h^6}{J_{\text{eff}}^5}$ for $i $ odd       \\ \hline
$u^m_i$                             & $\sim \frac{h^6}{J_{\text{eff}}^5}$ for $i $ even      \\ 
                                    & $ \sim \frac{h^4}{J_{\text{eff}}^2 J_2}$ for $i$ odd  \\ \hline
$t^{e,\parallel}_{ij}$ & $\sim \frac{h^5}{J^2_{\text{eff}} J^2_2}$ for $i$ even          \\ 
                                    & $\sim \frac{h^9}{J_{\text{eff}}^7J_1}$ for $i$ odd     \\ \hline
$t^{m,\parallel}_{ij}$ & $\sim \frac{h^9}{J_{\text{eff}}^7J_1}$ for $i$ even    \\ 
                                    & $\sim \frac{h^5}{J^2_{\text{eff}} J^2_2}$ for $i$ odd         \\ \hline
$t^{e,\perp}_{ij}$ & $\sim \frac{h^{15}}{J_{\text{eff}}^{13}J_1}$ for all $i$ \\ \hline
$t^{m,\perp}_{ij}$ & $\sim \frac{h^{15}}{J_{\text{eff}}^{13}J_1}$ for all $i$ \\ \hline
\end{tabular}

\caption{The perturbative values of the coefficients used in Eq. \ref{eq:loopHam0} for the  $(m,r) = (2,1)$ twist angle. The lattice site $i = (x,y)$ is even when $x +y = 0 \text{ mod}(2)$ and odd when $x +y = 1 \text{ mod}(2)$.}
\label{tab:Coeffs2}
\end{table}

\section{Subleading Corrections to the Defect Hamiltonian}\label{App:Sub}
Here we will consider subleading corrections to the effective Hamiltonian for the defect lattice Eq. \ref{eq:loopHam0} (equivalently Eq. \ref{eq:loopHamI0}). In our initial analysis, we only considered operators that create short $e$ and $m$ anyon loops. In this limit, the $ee$ and $em$ defects decouple. Here, we will consider subleading terms that couple the $ee$ and $em$ defects. We will also consider the effects of including terms that create $\psi$ anyon loops. To analyze these subleading contributions we will need to consider both the $ee$ and $em$ defects. To this end, we will define two new types of anyon loop creation operators. First we will define the operator $\bar{W}^{e(m)}_{k}$ that creates an $e(m)$ anyon loop on the top layer around the $em$ defect at site $k$. It is important to note that since the defect at site $k$ is an $em$ defect, an $e(m)$ anyon loop around the $k^{\text{th}}$ defect on the top layer is equivalent to an $m(e)$ anyon loop around the $k^{\text{th}}$ defect on the bottom layer. Second, there is the operator $\bar{W}^{e(m)}_{kl}$ that creates an $e(m)$ anyon string on the top layer that passes through the $em$ defect at site $k$ and becomes a $m(e)$ anyon string on the bottom layer and then returns to the top layer through the $em$ defect at site $l$ to form a closed loop. The leading order Hamiltonian for the $ee$ and $em$ defects is then given by 
\begin{equation}
\begin{split}
H_{\text{defect}} = &-\sum_{i\in ee} (u^e_{i} W^e_{i} + u^m_{i} W^m_{i})\\ &-\sum_{\langle i j\rangle \in ee} (t^{e}_{i j} W^{e}_{i j} + t^{m}_{i j} W^{m}_{i j})\\ &-\sum_{k \in em} (\bar{u}^e_{k} \bar{W}^e_{k} + \bar{u}^m_{k} \bar{W}^m_{k})\\ &-\sum_{\langle k l\rangle \in em} (\bar{t}^{e}_{k l} \bar{W}^{e}_{kl} + \bar{t}^{m}_{kl} \bar{W}^{m}_{kl}),
\label{eq:AloopHam0}
\end{split}
\end{equation}
where the first two sums are over the sites of the $ee$ defects, and the second two sums are over the sites of the $em$ defects. The operators $W^{e(m)}_{i}$ and $W^{e(m)}_{ij}$, and couplings $u^{e(m)}_{i}$ and $t^{e(m)}_{i j}$ are defined as in Eq. \ref{eq:loopHam0}. The coupling $\bar{u}^{e(m)}_{k}$ is the amplitude to create an anyon loop around the $k^{\text{th}}$ $em$ defect, and $\bar{t}^{e(m)}_{kl}$ is the amplitude to create an anyon loop the passes through the $k^{\text{th}}$ and $l^{\text{th}}$ $em$ defects. The non-trivial commutation relationships for the new anyon loop creation operators are $\{\bar{W}^e_k, \bar{W}^m_{kl} \} = \{\bar{W}^e_l, \bar{W}^m_{kl} \} = 0$, and $\{\bar{W}^m_k, \bar{W}^e_{kl} \} = \{\bar{W}^m_l, \bar{W}^e_{kl} \} = 0$. If we define two new sets of Pauli matrices $\bar{\tau}_1$ and $\bar{\tau}_2$, we can satisfy the new anyon loop algebra with the following identifications: $\bar{W}^e_{k} = \bar{\tau}^x_{1,k}$, $\bar{W}^m_{k} = \bar{\tau}^x_{2,k}$, $\bar{W}^m_{kl} = \bar{\tau}^z_{1,k}\bar{\tau}^z_{1,l}$ and $\bar{W}^e_{kl} = \bar{\tau}^z_{2,k}\bar{\tau}^z_{2,l}$. Using this, Eq. \ref{eq:AloopHam0} can be converted into four Ising models:
\begin{equation}
\begin{split}
    H_{\text{defect}} = &-\sum_{i\in ee} (u^e_{i} \tau^x_{1,i} +u^m_{i} \tau^x_{2,i})\\ &-\sum_{\langle i j\rangle \in ee} (t^{e}_{i j} \tau^z_{2,i}\tau^z_{2,j} + t^{m}_{i j} \tau^z_{1,i}\tau^z_{1,j} )\\&-\sum_{k \in em} (\bar{u}^e_{k} \bar{\tau}^x_{1,k} +\bar{u}^m_{k} \bar{\tau}^x_{2,k})\\ &-\sum_{\langle k l\rangle \in em} (\bar{t}^{e}_{kl} \bar{\tau}^z_{2,k}\bar{\tau}^z_{2,l} + \bar{t}^{m}_{kl} \bar{\tau}^z_{1,k}\tau^z_{1,l} ),
\label{eq:AloopHamI0}
\end{split}
\end{equation}
where $\tau_1$ and $\tau_2$ are defined as in Eq. \ref{eq:loopHamI0}. From our analysis in Appendix \ref{App:MIM}, we know that each Ising model will be in one of two phases: a symmetry broken ferromagnetic phase, and a trivial paramagnetic phase. For example, $\langle \tau^z_{1,i} \rangle = \pm 1$ when the $\tau_1$ Ising spins are in the ferromagnetic phase and $\langle \tau^x_{1,i} \rangle = 1$ when the $\tau_1$ Ising spins are in the paramagnetic phase. 

We will now discuss several types of subleading corrections to Eq. \ref{eq:AloopHamI0}. First, there are the contributions coming from processes that create an anyon loops that encircle both an $ee$ and an $em$ defect. An operator that creates such an anyon loop can be decomposed into the product of an operator that creates an anyon loop around an $ee$ defect, an operator that creates an anyon loop around an $em$ defect, and a trivial operator that creates a contractible loop that can be ignored.  
Second, there are the contributions coming from processes that create anyon loops that pass through both a pair of $ee$ defects and a pair of $em$ defects. An operator that creates such an anyon loop can be decomposed into the product of an operator that creates an anyon loop that passes through a pair of $ee$ defects, an operator that creates an anyon loop that passes through a pair of $em$ defects, and a trivial operator that creates a contractible anyon loop that can also be ignored.

The contributions to the effective Hamiltonian from these operators are (in terms of the Ising spins):
\begin{equation}
    \begin{split}
        H_{\text{int}} = &-\sum_{i \in ee}  \sum_{k \in em}\Big[ q^{11}_{ik} \tau^x_{1,i}\bar{\tau}^x_{1,k}+q^{12}_{ik} \tau^x_{1,i}\bar{\tau}^x_{2,k}\\ &+ q^{21}_{ik} \tau^x_{2,i}\bar{\tau}^x_{1,k}+q^{22}_{ik} \tau^x_{2,i}\bar{\tau}^x_{2,k}\Big]
        \\&- \sum_{\langle ij \rangle \in ee} \sum_{\langle kl \rangle \in em} \Big[ g^{11}_{ijkl} \tau^z_{1,i}\tau^z_{1,j}\bar{\tau}^z_{1,k}\bar{\tau}^z_{1,l}\\ &+ g^{12}_{ijkl} \tau^z_{1,i}\tau^z_{1,j}\bar{\tau}^z_{2,k}\bar{\tau}^z_{2,l}+ g^{21}_{ijkl} \tau^z_{2,i}\tau^z_{2,j}\bar{\tau}^z_{1,k}\bar{\tau}^z_{1,l}\\ &+ g^{22}_{ijkl} \tau^z_{2,i}\tau^z_{2,j}\bar{\tau}^z_{2,k}\bar{\tau}^z_{2,l}\Big].
    \end{split}
    \label{eq:ALoopHamInt1}
\end{equation}
Here, the $q$ couplings are the amplitudes to create a given anyon loop that encircles both an $ee$ and $em$ defect, and the $g$ couplings are the amplitudes to create a given anyon loop that passes through a pair of $ee$ defects and a pair of $em$ defects. Since these effects are subleading, we can consider their effect on the ground states of Eq. \ref{eq:AloopHamI0}. When the spins are in a ferromagnetic ground state, the $q$ coupling is irrelevant, since it flips two pairs of spins. The $g$ coupling in the ferromagnetic phase simply shifts the ground state energy, since $\langle \tau^z_{1/2,i}\tau^z_{1/2,j}\rangle = \langle \bar{\tau}^z_{1/2,k}\bar{\tau}^z_{1/2,l}\rangle = 1$. Similarly, in the paramangetic phase, the $g$ coupling is irrelevant, and the $q$ coupling shifts the ground state energy. We can thereby conclude that these subleading terms in Eq. \ref{eq:ALoopHamInt1} only shift the energy of the ground states of Eq. \ref{eq:AloopHamI0}. 

There are also additional subleading terms due to the $\psi$ anyon loops. Processes that create $\psi$ anyon loops were initially ignored since the $\psi$ anyons have a larger excitation gap in the Kitaev honeycomb model than the $e$ and $m$ anyons. Due to the fusion rule $e \times m = \psi$, the operators that create $\psi$ anyon loops can be decomposed into the product of an operator that creates an $e$ anyon loop and an operator that creates an $m$ anyon loop. For instance, the operator that creates a $\psi$ anyon loop around the $i^{\text{th}}$ defect can be written as $W^{\psi}_i \equiv W^e_i W^m_i$. As before, we are interested in two types of processes: those that create a $\psi$ anyon loop around a defect, and those that create a $\psi$ anyon loop that passes through a pair of defects. It is important to note that a $\psi$ anyon remains a $\psi$ anyon after passing through a $em$ defect, since $ \psi = e \times m \rightarrow m \times e = \psi$. This means that there is a finite amplitude to create a $\psi$ anyon loop that passes from one layer to another through an $ee$ defect, and returns to the first layer through an $em$ defect. We thereby have to consider processes where a $\psi$ anyon loop passes through a pair of $ee$ defects, a pair of $em$ defects, and a single $ee$ and a single $em$ defect. Including these process, the effective Hamiltonian for the $\psi$ anyon loops in terms of the Ising spins is
\begin{equation}
\begin{split}
    H_{\psi} = &-\sum_{i\in ee} u^\psi_{i} \tau^x_{1,i}\tau^x_{2,i}-\sum_{\langle i j\rangle \in ee} t^{\psi}_{i j} \tau^z_{2,i}\tau^z_{2,j}\tau^z_{1,i}\tau^z_{1,j} \\&-\sum_{k \in em} \bar{u}^{\psi}_{k} \bar{\tau}^x_{1,k}\bar{\tau}^x_{2,k}-\sum_{\langle k l\rangle \in em} \bar{t}^{\psi}_{kl} \bar{\tau}^z_{2,k}\bar{\tau}^z_{2,l}\bar{\tau}^z_{1,k}\tau^z_{1,l}\\ &-\sum'_{\langle i k\rangle} \tilde{t}^{\psi}_{ik} {\tau}^z_{1,i}{\tau}^z_{2,i}\bar{\tau}^z_{1,k}\tau^z_{2,k}.
\label{eq:AloopHamIP}
\end{split}
\end{equation}
The coupling $u^\psi_i$ ($\bar{u}^\psi_k$) creates a $\psi$ anyon loop around the $i^\text{th}$ ($k^{\text{th}}$) $ee$ ($em$) defect, and $t^{\psi}_{i j}$ ($\bar{t}^{\psi}_{k l}$) creates a $\psi$ anyon loop that passes through the $i$ and $j$ ($k$ and $l$) $ee$ ($em$) defects. The primed sum is over neighboring defect sites $i$ and $k$, where $i$ is an $ee$ defect, and $k$ is an $em$ defect. The coupling $\tilde{t}^{\psi}_{ik}$ is the amplitude to create a $\psi$ anyon loop that passes through the $i^{\text{th}}$ $ee$ defect and the $k^{\text{th}}$ $em$ defect.

The $u^\psi_i$ and $t^{\psi}_{i j}$ couplings turn the two original decoupled $\tau_1$ and $\tau_2$ Ising models in Eq. \ref{eq:AloopHamI0} into a 2D quantum Askin Teller model\cite{ashkin1943}. Similarly, the $\bar{u}^\psi_i$ and $\bar{t}^{\psi}_{i j}$ couplings turn the two original decoupled $\bar{\tau}_1$ and $\bar{\tau}_2$ Ising models into a second 2D quantum Ashkin-Teller model. We will now consider the effects of these weak Ashkin-Teller couplings on the ground state of Eq. \ref{eq:AloopHamI0}. When the $\tau_1$ and $\tau_2$ spins are in the paramagnetic phase, the $t^{\psi}_{i j}$ coupling is irrelevant, and the $u^\psi_i$ coupling shifts the ground state energy. When the $\tau_1$ and $\tau_2$ spins are in the ferromagnetic phase, the $u^\psi_i$ coupling is irrelevant, and the $t^{\psi}_{i j}$ coupling shifts the ground state energy. So the $u^\psi_i$ and $t^{\psi}_{i j}$ terms simply change the ground state energy of the Ising models. The analysis for $\bar{u}^\psi_i$, and $\bar{t}^{\psi}_{i j}$ is identical. The contributions from the $ \tilde{t}^{\psi}_{ik}$ coupling, however, are non-trivial when the Ising spins are in the ferromagnetic phase. To see this, we note that in the ferromangetic phase of Eq. \ref{eq:AloopHamI0}, $\langle \tau^z_{1,i}\rangle = \pm 1$, $\langle \tau^z_{2,i}\rangle = \pm 1$, $\langle \bar{\tau}^z_{1,k}\rangle = \pm 1$, $\langle \bar{\tau}^z_{1,k}\rangle = \pm 1$, leading to 16 ground states. The $ \tilde{t}^{\psi}_{ik}$ term in Eq. \ref{eq:AloopHamIP} lowers the energy of ground state configurations where $\langle \tau^z_{1,i}\tau^z_{2,i}\bar{\tau}^z_{1,k}\bar{\tau}^z_{2,k}\rangle = 1$. It can readily be verified that this reduces the number of ground states from $16$ to $8$. In the paramagnetic region, the $ \tilde{t}^{\psi}_{ik}$ term is irrelevant. 

\end{appendix}

\bibliography{Draft1120Notes.bib}
\bibliographystyle{apsrev4-1}

\end{document}